\pdfoutput=1

\documentclass[12pt,letterpaper]{article}
\usepackage{jheppub}
\usepackage{journals}
\usepackage{caption}
\usepackage{graphicx}
\usepackage{amssymb}
\usepackage{placeins}
\usepackage{jheppub}
\usepackage{journals}
\usepackage{graphicx}
\usepackage{amsmath}
\usepackage{placeins}
\usepackage{wrapfig}
\usepackage{subfig}
\usepackage{amssymb}
\usepackage{hyperref}
\usepackage[toc,page]{appendix}
\usepackage{tikz}

\newcommand{\SW}[1]{{\textcolor{red}{ [SW:#1]}}}
\newcommand{\SWn}[1]{{\textcolor{brown}{ [$SW^2$:#1]}}}
\newcommand{\LY}[1]{{\textcolor{blue}{ [LY:#1]}}}
\renewcommand{\SW}[1]{#1}
\renewcommand{\LY}[1]{#1}
\renewcommand{\SWn}[1]{#1}
\newcommand\Nfour{$\mathcal N\,{=}\,4$}

\begin		{document}

\title		{Colliding localized, lumpy  holographic shocks with a granular nuclear structure}

\author[a]{Sebastian Waeber,}
\author[b]{Laurence G.~Yaffe}

\affiliation[a] {Department of Physics, Technion, Haifa 32000, Israel
}
\affiliation[b] {Department of Physics, University of Washington, Seattle WA 98195-1560, USA
}
\emailAdd	{wsebastian@campus.technion.ac.il} \emailAdd	{yaffe@phys.washington.edu}

\abstract
    {%
    We apply a recent and simple technique which  speeds up the
    calculation of localized collisions in holography to study  more
    realistic models of the pre-hydrodynamic phase of heavy ion collisions
    using gauge/gravity duality.
    Our initial data reflects the lumpy nuclear structure of
    real heavy ions and our projectiles' aspect ratio mimics
    the Lorentz contraction of nuclei during RHIC collisions.
    At the hydrodynamization time of the central region of the quark
    gluon plasma developed during the collision, we find that most  of
    the system's vorticity is located well outside
    the hydrodynamized part of the plasma.
    Only the relativistic corrections to the thermal
    vorticity within the hydrodynamized region are non-negligible.
    We compare the transverse flow shortly after the collision 
    with previous results which did not use granular initial conditions and
    determine the proper energy density and fluid velocity in the
    hydrodynamized subregion of the plasma.
    }
\maketitle
\section{Introduction}
\label{intro}
Numerical calculations of holographic models of heavy ion collisions,
via high-accuracy solutions of five-dimensional Einstein equations \cite{Che},
require very substantial calculational resources in both run-time
and memory if the initial data is chosen to closely mimic the energy
density of incoming nuclei in heavy ion collisions at, e.g., RHIC.
Holographic calculations to date, despite using initial data modeling
simplified and rather unrealistic descriptions of real nuclei,
have yielded insight into significant aspects of the early phase
of heavy ion collisions including the onset of hydrodynamic behavior,
the domain of validity of hydrodynamic descriptions,
pre-hydro development of radial and transverse flow,
near-universal rapidity dependence, and more \SWn{\cite{che3,Chesler:2010bi,Che,Chesler:2015fpa,Casalderrey-Solana:2013aba,wae3,1507.08195,1307.2539,1607.05273}}.
However, many interesting questions involving the early phase
of quark-gluon plasma dynamics remain unexplored,
impeded by the computational challenges involved in solving
5D Einstein equations in geometries with no dimensionality-reducing
symmetries and with spatio-temporal structure whose accurate representation
requires a very large dynamic range.
Some of these questions, not yet adequately explored, include the
effect of initial state fluctuations on the formation and early stage
dynamics of produced quark-gluon plasma, the interplay between energy density
fluctuations and dependence on charge and flavor densities,
and the evolution of plasma vorticity, 
as well as the effect of finite 't Hooft coupling corrections 
needed to more closely model real QCD.

In this work we focus on enabling holographic modeling of
early stage heavy ion collisions with initial data which
closely mimics the granular structure of real nuclei.
The underlying dual field theory is the strong coupling limit
of maximally supsersymmetric Yang-Mills theory (\Nfour\ SYM),
not real QCD for which no correct dual holographic description is known.
In other words, we are approximating the dynamics of quark-gluon plasma (QGP)
produced in real heavy ion collisions,
a highly relativistic and strongly coupled non-Abelian plasma,
by the dynamics of \Nfour\ SYM plasma in its strong coupling
(and large $N_{\rm c}$) limit.
This, to be sure, is a drastic approximation.
As \Nfour\ SYM is a conformal theory, unlike QCD,
trying to model QCD using \Nfour\ SYM completely eliminates
all dynamics related to hadronization and actual particle production.
At best, holographic models based on \Nfour\ SYM can mimic the
behavior of real QGP during early stages of a collision where
the quark-gluon plasma does behave like a near-conformal fluid.%
\footnote
    {%
    There are non-conformal theories with known holographic descriptions
    some of which, while still differing from QCD, might be suitable for
    providing more controlled models of hadronization.
    Addressing such late-stage dynamics is outside the scope of the present work.
    }

In assessing the utility of holographic modeling of heavy ion collisions,
one should bear in mind that available alternative treatments for modeling
early stage dynamics in these collisions make at least equally large approximations.
Many studies have used a Glauber model of the initial projectile energy densities
directly as hydrodynamic initial data, as if there were no non-trivial
pre-hydrodynanmic evolution whatsoever \SWn{\cite{phobos,phobos2}}.
Much effort has also been devoted to studying \emph{asymptotically}
high energy collisions, leading to the development of the
Color Glass Condensate (CGC) description of collisions which may
be viewed as involving high occupancy of very weakly coupled partons \SWn{\cite{cgc}}.
This asymptotic regime, with a plethora of scales differing by powers of the
weak coupling, is far from what is achievable in 
experimentally accessible collisions.
Modeling which uses a CGC-inspired treatment of the initial state
to generate initial data for hydrodynamic evolution amounts to
converting, instantly, from an asymptotically weakly coupled description
to a near-ideal fluid description in which microscopic constituents
are strongly interacting and correlation lengths
are shorter than any other relevant scale.
This is intrinsically inconsistent, but reflects the reality that
there are no fully controlled calculational techniques for studying
the dynamics of real QGP as produced in current experiments.

Holographic modeling based on \Nfour\ SYM provides a description of
early stage dynamics which incorporates, correctly,
the strong-coupling dynamics of a not-quite QCD non-Abelian plasma.
The resulting treatment is complementary to CGC-inspired models that
involve extrapolations of asymptotically weak coupling descriptions
to experimentally accessible collisions
in which the produced plasma is not weakly coupled.
For the remainder of this paper, we take as given this motivation
for using holographic modeling based on \Nfour\ SYM to study early
stages of relativistic heavy ion collisions.%
\footnote
    {%
    There are, of course, important probes of heavy ion collisions
    involving high transverse momentum jets and produced particles
    for which holographic modeling is not appropriate.
    The goal of holographic modeling is to capture the
    dynamics of the bulk of the produced plasma, not
    high momentum tails of distributions for which the
    asymptotic freedom of QCD is essential.
    }

While the earlier holographic calculation in \cite{Che} captured qualitative
features of a collision of projectiles somewhat resembling colliding nuclei,
the aspect ratios of the projectiles
considered in \cite{Che} were an order of magnitude smaller than the aspect
ratios of (lab frame) Lorentz contracted nuclei in RHIC collisions.
The resource requirements (in both  run time and memory) of the
most demanding steps in computing these collision,
without resorting to any computational approximations,
increase approximately quadratically with increasing aspect ratios.
However, if one hopes to make quantitative statements about observables
that are sensitive to the ratio between the transverse and longitudinal
scales, such as the vorticity, it is necessary to work with projectiles
with realistic Lorentz contractions.
Moreover, there is compelling evidence that transverse fluctuations in the
energy densities of colliding nuclei have large influence on the resulting
plasma evolution and, in particular, that strong fluctuations are necessary
to account for the size of odd azimuthal flow moments $v_{2\,i +1}$ observed
in experiments \cite{bayesian,Alice}.
These flow moments $\{ v_n \}$ are the Fourier expansion coefficients
(in azimuthal angle)
of the transverse plane particle distribution,
\begin{equation}
    E \, \frac{d^3 N}{dp^3}
    = \frac{1}{2\pi} \,
    \frac{d^2N}{p_T dp_T dy} \,
    \Big(1+2 \sum_{n=0}^\infty v_n \cos\big(n(\phi-\Psi_n)\big) \Big),
\end{equation}
with $E$ the energy, $p$ momentum, $p_T$ transverse momentum,
$\phi$ the azimuthal angle, $y$ the pseudorapidity of a final state
particle, and $\Psi_n$ the $n$-th harmonic symmetry plane angle \SWn{\cite{9407282}}.
The observation of
large odd moments, which would be suppressed if the overlap region of the
projectiles during the collision was perfectly smooth,
imply strong transverse fluctuations \cite{Alice}.

In the present work, our goal is to demonstrate the feasibility of
computing holographic collisions with initial data modeling far more
realistic collisions than has previously been possible, and examine
the resulting implications for the onset of hydrodynamic behavior
as well as the development of pre-hydrodynamic flow and vorticity.
In particular, we will incorporate initial state fluctuations
in energy density along the lines of the treatment in
\SWn{\cite{phobos,phobos2}}, and
an aspect ratio of our projectiles which matches the Lorentz contraction
of RHIC collisions.

Attempting to perform this calculation using exactly the same calculational
techniques employed in \cite{Che},
involving a characteristic formulation of Einstein's equations,
spectral approximations for the resulting partial differential equations,
and relying a sufficiently large non-distributed unitary memory system,
would not be feasible -- at least on systems to which we have full-time access.
To make this calculation feasible, we will employ the transverse derivative
expansion procedure developed in \cite{2206.01819}.
As shown in that work, expanding in transverse derivatives produces
a simple, yet effective technique for computing approximate but quite accurate
solutions to localized holographic collisions.
By expanding in transverse gradients up to first order in derivatives
we could reproduce the exact solutions, for intervals up to the hydrodynamization time,
to within errors in the range of 1-10\%,
using only a small fraction of the run-time and memory that would be needed
for the exact calculation with no expansion in transverse gradients.

We will apply this technique to compute, via holography, the collision
of projectiles with a lumpy, granular structure, reflecting the
nuclear structure of heavy ions.
For the initial data we use a Lorentz-contracted Woods-Saxon potential as
the probability distribution of the centers of the individual nucleons.
The Lorentz contraction factor will reflect energies at RHIC collisions.
We enforce a minimal distance of the nucleons' centers to ensure limited
overlap as in \cite{phobos, phobos2}.
The nuclear model giving rise to our holographic initial data also takes
into account a realistic skin thickness of the nuclei.

\SW{
We will find that the time at which \LY{ roughly} half of the central, low rapidity  region can be described by hydrodynamics  approximately corresponds 
to the hydrodynamization time of the same region observed during collisions of smooth Gaussians without a lumpy structure. This is in line with the expectations from  \cite{mue}, which predicted that granular initial data should delay full hydrodynamization by about a factor of 2.}
The vorticity, at the time when the majority of the central region of the quark gluon plasma has hydrodynamized,
is dominated by contributions far away from the central region, with
\SW{only a small fraction of the vorticity} in the system deposited in the hydrodynamized center.
The calculation presented in this work is a natural extension of the model discussed in
\cite{Chesler:2015fpa}, where the authors approximated heavy ions by smooth Woods-Saxon
potentials, studied central collisions via planar shockwave collisions in holography,
and only included transverse dynamics later on in the hydrodynamic evolution.

\section{Initial data and nuclear model}
\label{nuclear_model}

Following \cite{che3, Chesler:2010bi,Che},
we first formulate the metric for a single shockwave in AdS$_5$
using Fefferman-Graham coordinates,
\begin{equation}
ds_{FG}^2= \frac{1}{\rho^2}\big(-dt^2+d\rho^2+(d\bold x^\bot)^2+dz^2+\rho^4 h_{\pm}(\bold x^\bot,z^\mp,\rho) (dz^\pm)^2 \big) \,,
\label{FG}
\end{equation}
with $z^\mp = z \mp t$, and $\rho$ an inverted radial coordinate.
The Einstein equations require 
\begin{equation}
\Big(\frac{d^2}{d \rho^2}  -\frac{3}{\rho} \frac{d}{d \rho} +\,\nabla_\bot^2\Big) \rho^4\, h_\pm =0 \,.
\label{Ein_init}
\end{equation}
In the dual quantum field theory,
the metric (\ref{FG}) corresponds to a state  with
\begin{subequations}
\begin{align}
\langle T^{00} \rangle &=\langle T^{zz} \rangle = \frac{N_c^2}{2 \pi^2} \,
h_{\pm}\Big|_{\rho=0} \,,
\\
\langle T^{0z} \rangle &= \pm \frac{N_c^2}{2 \pi^2} \,
h_{\pm}\Big|_{\rho=0}.
\end{align}
\end{subequations}
Due to the large aspect ratios of the Lorentz contracted projectiles,
longitudinal gradients are much larger than transverse spatial gradients.
To simplify the problem
we exploit this separation of scales
by systematically expanding the Einstein equations in transverse derivatives.
We use the symbol $\mathcal{O}(\nabla_\bot^i)$ to represent terms
that are at least of $i$-th order in transverse derivatives.
(This is explained in more detail in the Appendix.)

Through first order in transverse derivatives,
the single shock function $h_{\pm}$ has no radial dependence
\begin{equation}
h_{\pm}(\bold x^\bot,z^\mp,\rho) = h_{\pm}(\bold x^\bot,z^\mp)+\mathcal{O}(\nabla_\bot^2).
\end{equation}
Otherwise (\ref{Ein_init}) does not constrain $ h_{\pm}(\bold
x^\bot,z^\mp)$  as a function of boundary coordinates, so it may
be chosen to be an arbitrary function of $\bold x^\bot$ and $z^\mp$.
We aim to choose $h_{\pm}$ so that the initial boundary stress energy tensor corresponds to a realistic model for a boosted gold nucleus.
The model we use is motivated by the standard model for heavy ions usually applied in Glauber Monte Carlo simulations \cite{phobos,phobos2}.
There the position of each nucleon in the nucleus is determined from a probability density function that can be thought of as the single-particle probability density in a quantum mechanical model.
We take this probability density to be a boosted spherically symmetric
distribution.
The radial distribution is derived from low energy electron
scattering experiments \cite{vries} and is given by a boosted Fermi
distribution with three shape parameters:
the nuclear radius $R$, the skin thickness $a$,
and the boost factor $\gamma$.
The resulting
probability distribution for the position of a nucleon
is a standard  Woods-Saxon potential,
\begin{equation}
P (\bold x^\bot,z^\mp) =  \frac{n}{1+\exp{\Big(\big(\sqrt{\bold (x^\bot)^2+\gamma^2 (z^\mp)^2}-R}\big)/a\Big)}.
\label{probability_distribution}
\end{equation}
The normalization constant $n$ is chosen such that $\int dx^3 \> P = 1$.
To model RHIC collisions, we use $\gamma = 100$ as the
longitudinal Lorentz contraction factor of each colliding nucleus.
The energy density of 
each nucleon is modeled as a Lorentz-contracted  Gaussian profile,
\begin{equation}
G_\pm(\bold x^\bot,z^\mp,\bold x^\bot_0,z^\mp_0)=
\frac{\mu^3}{\sqrt{2 \pi w^2/\gamma^2}}
\exp\big({}-\frac{ \gamma^2 }{2}(z^\mp{-}z^\mp_0)^2/w^2  \big)
\exp\big({}-\frac{1}{2} (\bold x^\bot {-}\bold x^\bot_0)^2/w^2\big),
\label{Gpm}
\end{equation}
centered around $(\bold x^\bot_0,z^\mp_0)$,
with the same Lorentz-contraction as in the nucleon distribution
(\ref{probability_distribution}).
To ensure that the individual nucleons have limited overlap,
we follow \cite{phobos, phobos2} and implement a minimal distance
$d_{\text{min}}$ between them. We do so by generating the ensemble
of nucleon centers in the following way: after choosing the $i$-th
nucleon center point $(\bold x^\bot_i,z^\mp_i)$, we update the
probability distribution (\ref{probability_distribution}) via
\begin{equation}
P \rightarrow P \times\Theta(|\bold x^\bot-\bold x^\bot_i|^2+\gamma^2(z^\mp-z^\mp_i)^2-d_{\text{min}}^2),
\end{equation}
with $\Theta$ a unit step function.
We repeat this procedure after each chosen nucleon center.
The projectile energy density function $h_{\pm}$ is then given
by the superposition
\begin{equation}
h_{\pm}(\bold x^\bot,z^\mp)= \sum_{i=0}^{196} G_\pm(\bold x^\bot,z^\mp,\bold x^\bot_i,z^\mp_i).
\label{hpm}
\end{equation}
Since we aim to simulate heavy ion collisions with realistic
parameters, we choose (as in, e.g., \cite{Chesler:2015fpa})
the scale $\mu$ determining the
the amplitude of $G_{\pm}$ such that
\begin{equation}
\frac{N_A \times 200 \, \text{GeV}}{2}
= E_{\text{RHIC}}= \int d^2 \bold x_{\bot} \> dz \; \langle T^{00} \rangle
= \frac{N_c^2}{2 \pi^2}\int d^2 \bold x_{\bot} \> dz \; h_{\pm}\Big|_{\rho=0},
\label{condition}
\end{equation}
with $N_A=197$ being the number of nucleons in a gold nucleus and $N_c=3$ the gauge group rank of QCD.
After choosing the skin thickness of the potential $a$, the minimal
distance $d_{\text{min}}$, the transverse size of  each nucleon $w$,
and the transverse size  $R$  of the probability distribution
(\ref{probability_distribution}) in units of $[1/\mu]$, the condition
(\ref{condition}) then fixes the amplitude $\mu $ in  (\ref{Gpm})
and allows us to give $R$, $a$, $d_{\text{min}}$, and $w$ in
units of $[1/\text{GeV}]$.

 We work with a nuclear model using the following parameters. As
 in \cite{Chesler:2015fpa} the transverse size $R$ of our probability
 distribution (\ref{probability_distribution}) is $6.5$ fm, and
 the skin thickness is set to $ 0.66$ fm; these values are close to nuclear
 parameters obtained from elastic electron scattering \cite{vries2}.
 The  minimal distance $d_{\text{min}}= 0.4$ fm, as in
 \cite{phobos2}, and each nucleon has transverse size $w = 1$ fm.
 These parameters lead to the value $\mu = 1.1$ GeV. Our nucleon
 size $w$ is larger than the typical nucleon size of $w \approx 0.5$
 fm argued for in \cite{schenke}. We use a somewhat larger nucleon size
 since it decreases the required longitudinal and transverse resolution and
 speeds up the computation. It should be noted that since our
 nucleons themselves are Gaussian energy density distributions, the
 actual skin thickness of the heavy ion model and the skin thickness
 $a$ of the probability distribution (\ref{probability_distribution})
 are not identical. Therefore we compute multiple ensembles of initial
 data following the above procedure, for various choices of $a$
 in (\ref{probability_distribution}). We then select the ensemble
 whose average possesses an actual skin thickness of $0.66$ fm and
 select two random samples from this ensemble, corresponding to
 left and right moving shocks. With our choice for $\mu$ the parameter
 $a$ in (\ref{probability_distribution}) is $0.165$ fm.

For the numerical evolution we work in units such that the longitudinally integrated energy density profile  of a single nucleus at vanishing transverse radius is normalized to one,
\begin{equation}
\int dz \> h_{\pm}(\bold x^\bot{=}0) =1.
\end{equation}
We then use the above parameter values to present results in physical units.

To construct initial data for the time evolution in a coordinate system in which one can employ the characteristic formulation of general relativity,
it is necessary to transform the metric ansatz (\ref{FG}) on the initial time slice  from Fefferman-Graham coordinates to infalling Eddington-Finkelstein coordinates, for which the metric has the form
 \begin{equation}
 ds_{EF}^2 = u^{-2} \Big(g^{EF}_{\mu \nu}(x,r) \> dx^\mu dx^\nu -2 \, dr du\Big).
 \label{EF}
 \end{equation}
We perform this transformation
order by order in transverse derivatives
following the method outlined in \cite{che3, Chesler:2010bi,Che,2206.01819}.
To compute the  coordinate transformation numerically we discretize spacetime
and use Fourier grids in spatial directions with $N_x=N_y=40$ and $N_z=256$
grid points, and a Chebyshev grid in the radial direction with three domains
and $N_u=3 \times 28$ grid points in total.
In \cite{2206.01819}, we show in detail  how to construct initial data as an expansion in transverse derivatives.
We choose an impact parameter $\vec b$  along the $x$ direction with
$|\vec b|=4.5$ fm.

\section{Time evolution}

To compute time evolution we expand the Einstein equations in
transverse derivatives and solve them order by order 
on each time slice through first order in transverse gradients.
We briefly review the main idea behind the transverse derivative
expansion in the Appendix.
A more thorough discussion of this expansion technique
and how to efficiently solve the transverse derivative expanded
Einstein equations may be found in \cite{2206.01819}. As shown there,
the approximation by a truncated expansion in transverse gradients
for collisions of shocks with large aspect ratios provides substantial
run time and memory improvements, while errors are $\lesssim 10 \%$
at the hydrodynamization time. On each time slice one has to solve
an elliptic partial differential equation to ensure that
the radial position of the  horizon remains stationary \cite{che3}.
By expanding in transverse gradients,
this equation simplifies from an elliptic
differential equation to a collection of ordinary differential
equations in the longitudinal coordinate.
This simplification is a major contributor
to the above-mentioned calculational improvements.

To solve the expanded Einstein equations numerically,
we use a two-domain Chebyshev grid with $N_u =2 \times 21 $ grid
points in radial direction of the AdS space and Fourier grids with
$N_x=N_y=40$ and $N_z=256$ in spatial directions.
Time evolution of the geometry is performed using a fourth order
Runge-Kutta algorithm with a physical time step size of
$\delta t=7.5 \times 10^{-4}$ fm/$c$.
Just using a Mathematica implementation running on a decade-old
multi-core desktop computer \LY{ with 128 Gb of memory},
we solve the geometry from $t_0= -0.144$ fm/$c$ to $t_1=0.144$ fm/$c$
in about three weeks of run time.%
\footnote
    {
    \LY{On newer machines with the same total memory,
    our codes run approximately twice as fast.}
    }
The initial projectiles have coinciding longitudinal positions at
time $t = 0$.
The initial time $t_0 = -0.144$ fm/$c$ 
is chosen such that the single shock bulk
solutions in the dual gravity theory do not overlap within the integration
domain.
The integration domain stretches between the boundary and
the apparent horizon of a smooth Schwarzschild black brane
which is always present in the two-shock geometry.
To improve numerical stability, we add a small uniform background
energy density equal to $7\%$  of the peak energy density of the individual
projectiles.
This background energy density has only minimal
influence on the evolution during the period we study.

\section{Results}
\subsection{Boundary stress energy tensor}

\begin{figure}
\begin{minipage}{0.5\textwidth}
\includegraphics[scale=0.5]{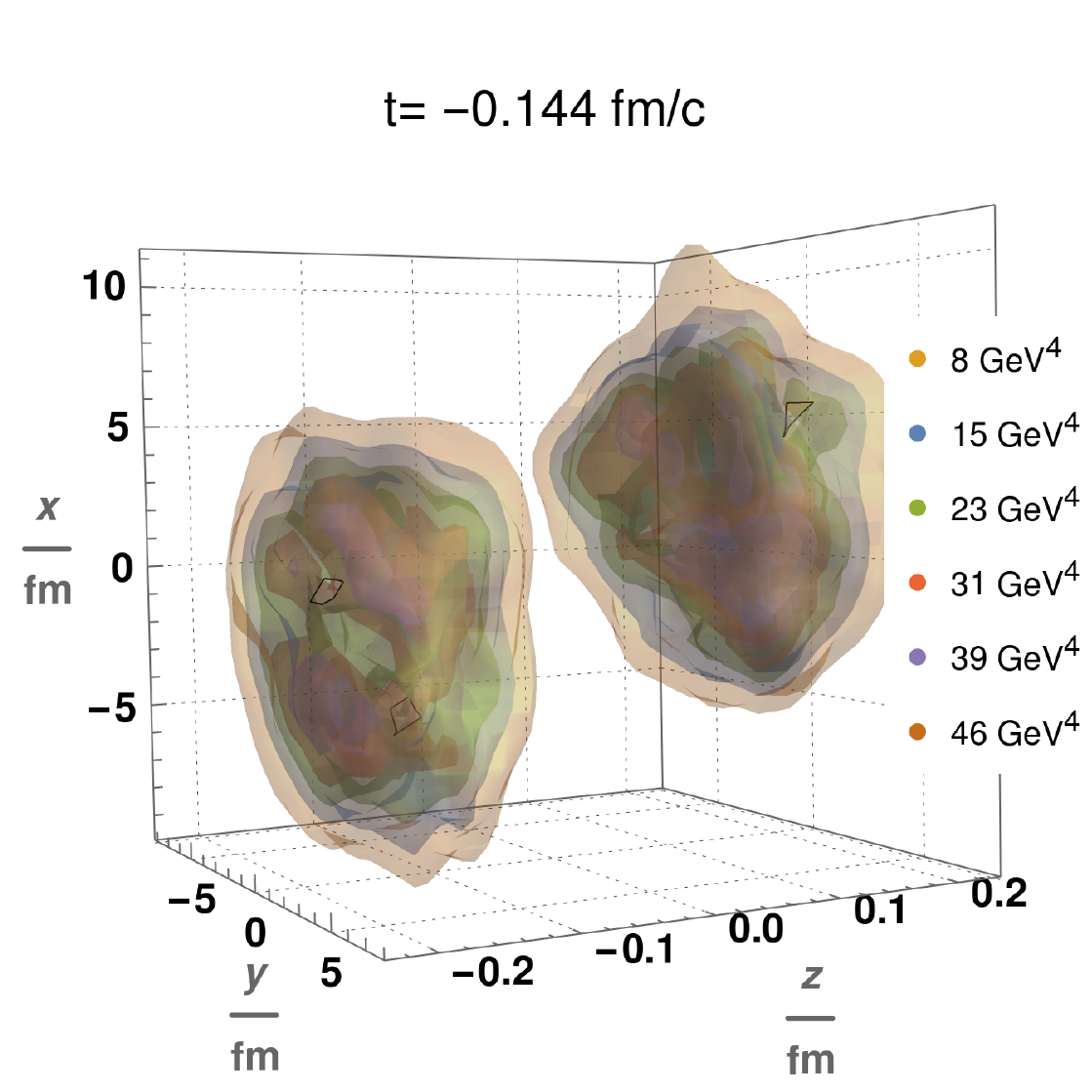}
\end{minipage}
\begin{minipage}{0.5\textwidth}
\includegraphics[scale=0.5]{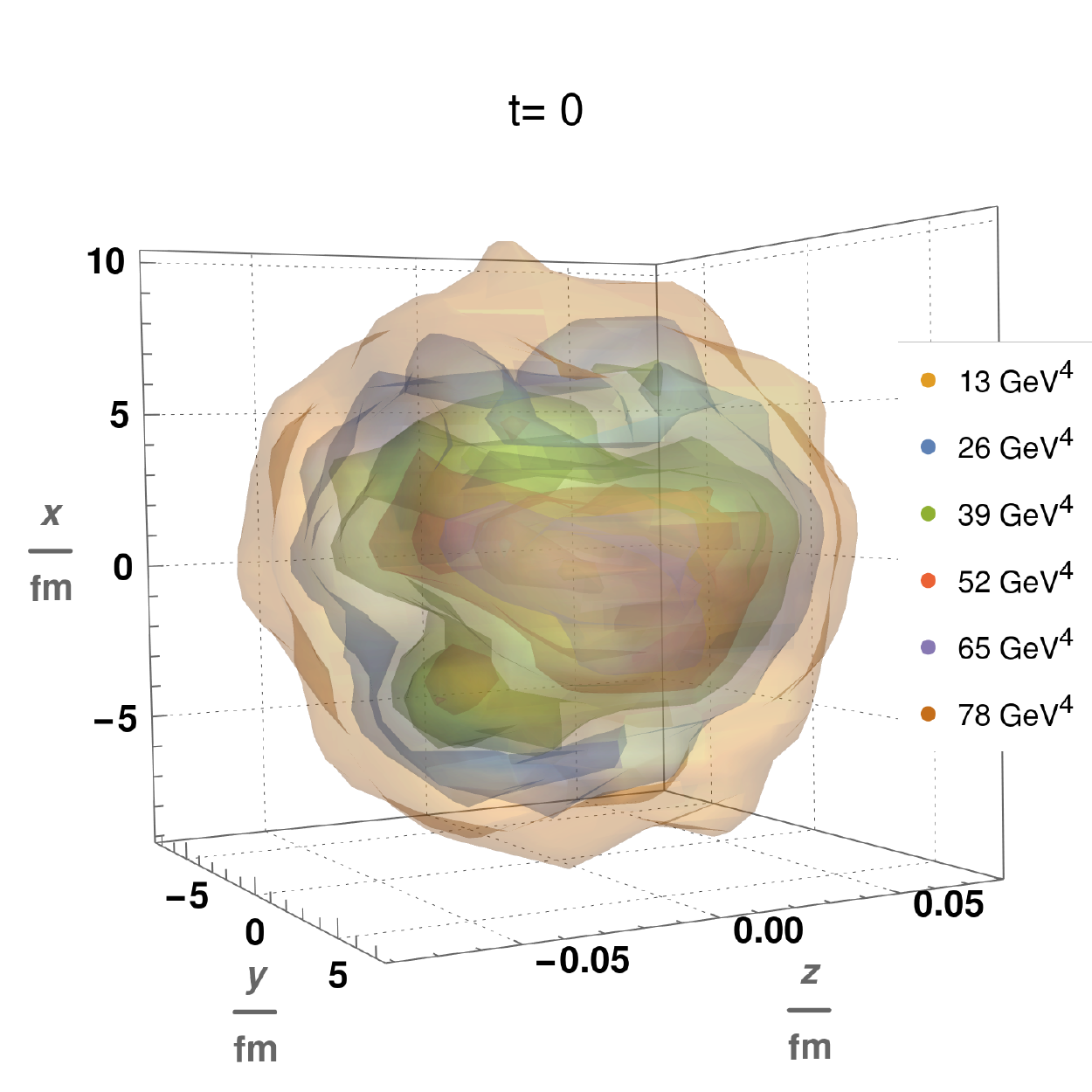}
\end{minipage}
\\\\\\
\begin{minipage}{0.5\textwidth}
\includegraphics[scale=0.5]{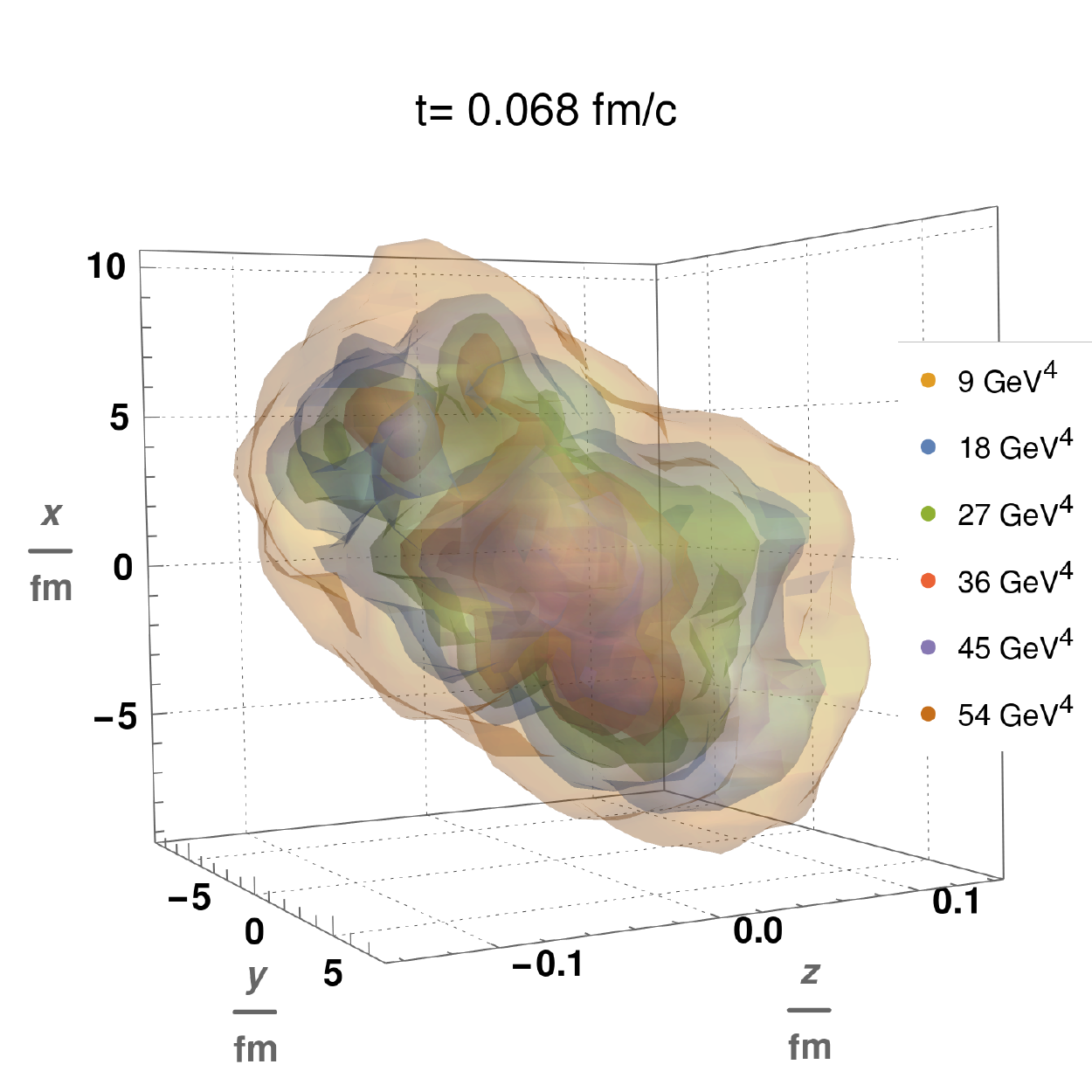}
\end{minipage}
\begin{minipage}{0.5\textwidth}
\includegraphics[scale=0.5]{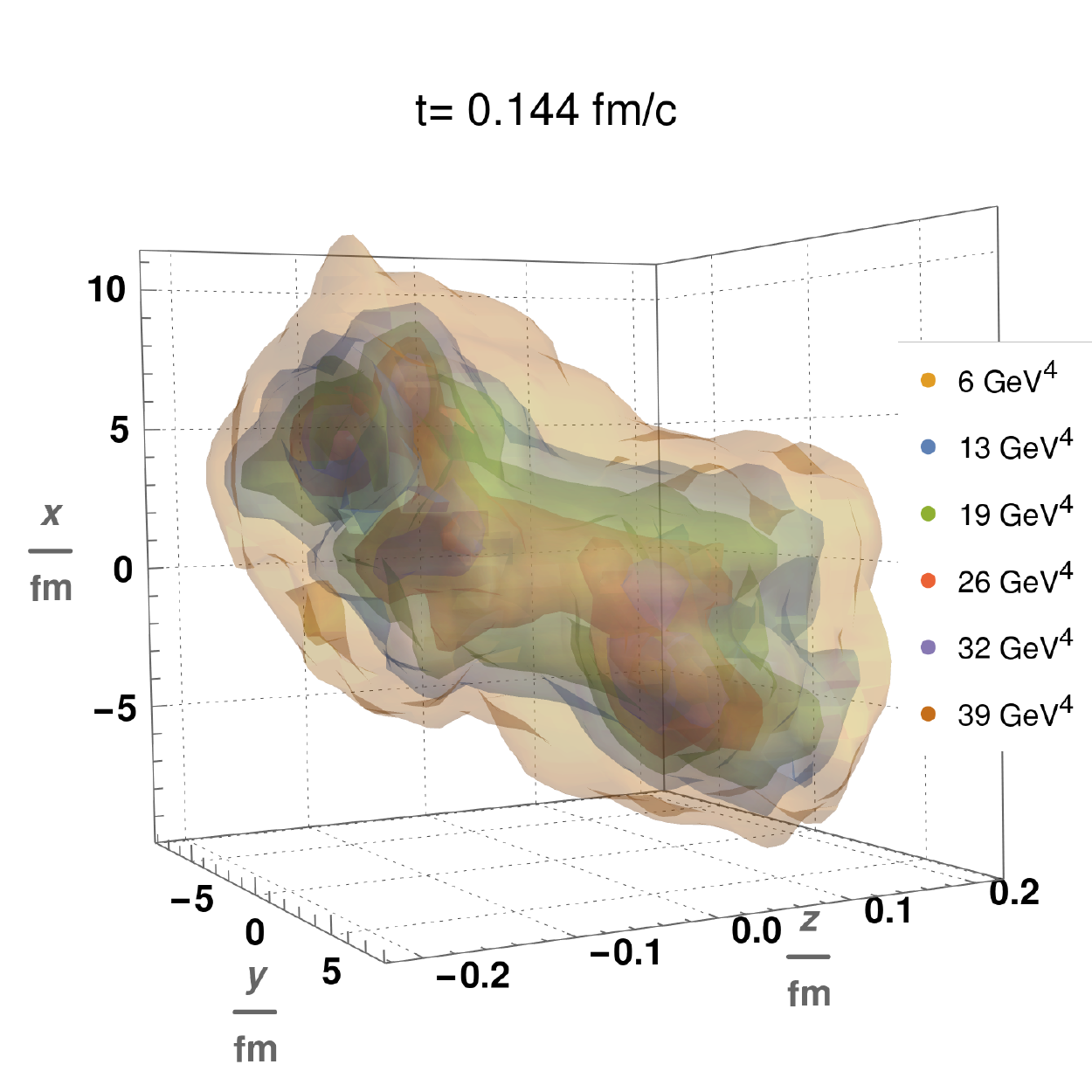}
\end{minipage}
\caption{The energy density distribution during the collision of localized, granular, highly contracted holographic shocks computed up to first order in transverse derivatives. The spatial directions are labeled in units of [fm]. From top left to bottom right: the surface plots of the energy density  evaluated at times $t= -0.144$ fm/$c$,  $t= 0$ fm/$c$,  $t= 0.068$ fm/$c$, and $t= 0.144$ fm/$c$. }
\label{contour_energy}
\end{figure}

\begin{figure}
\begin{minipage}{0.5\textwidth}
\includegraphics[scale=0.6]{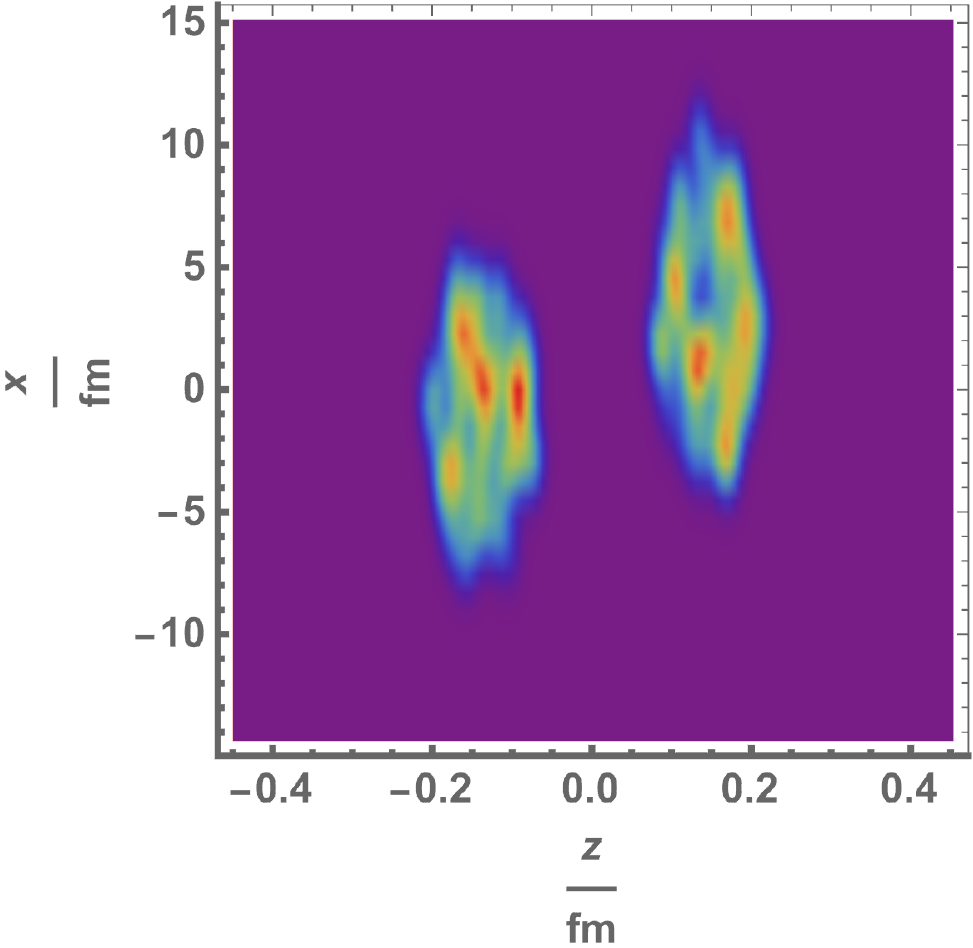}
\includegraphics[scale=0.6]{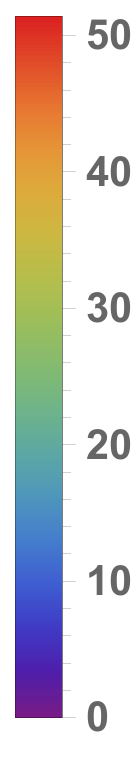}
\end{minipage}
\begin{minipage}{0.5\textwidth}
\includegraphics[scale=0.6]{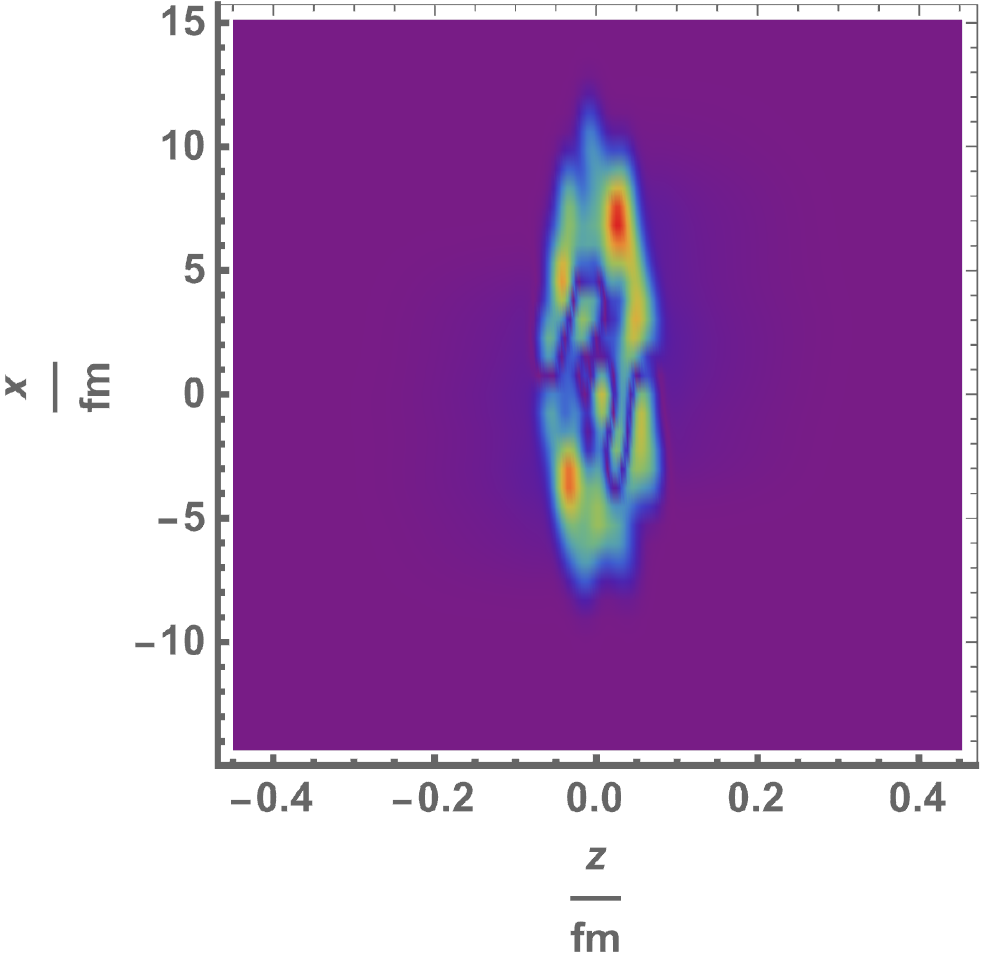}
\includegraphics[scale=0.6]{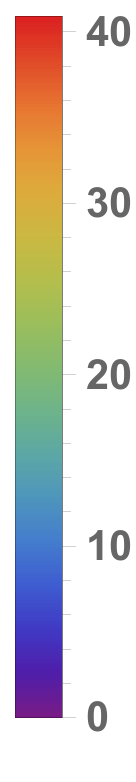}
\end{minipage}
\\\\\\
\begin{minipage}{0.5\textwidth}
\includegraphics[scale=0.6]{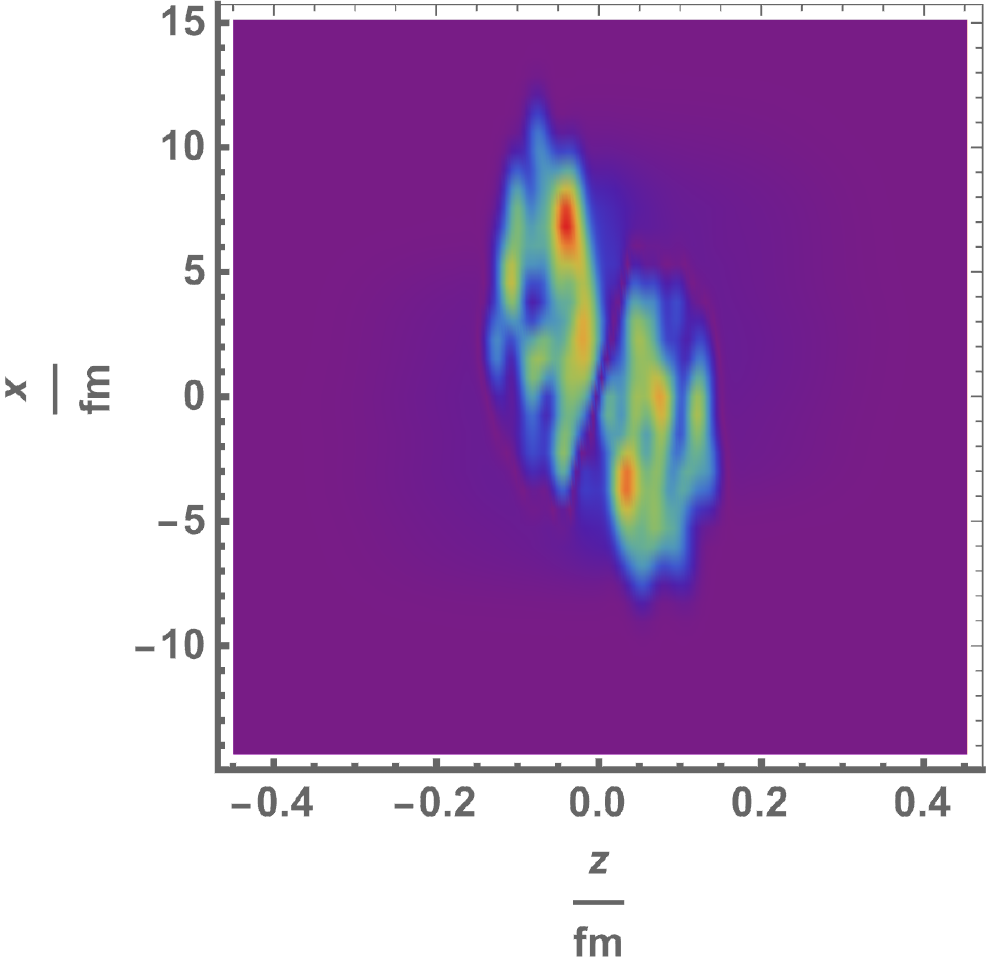}
\includegraphics[scale=0.6]{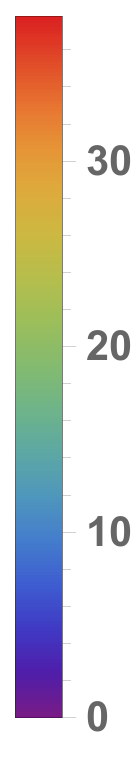}
\end{minipage}
\begin{minipage}{0.5\textwidth}
\includegraphics[scale=0.6]{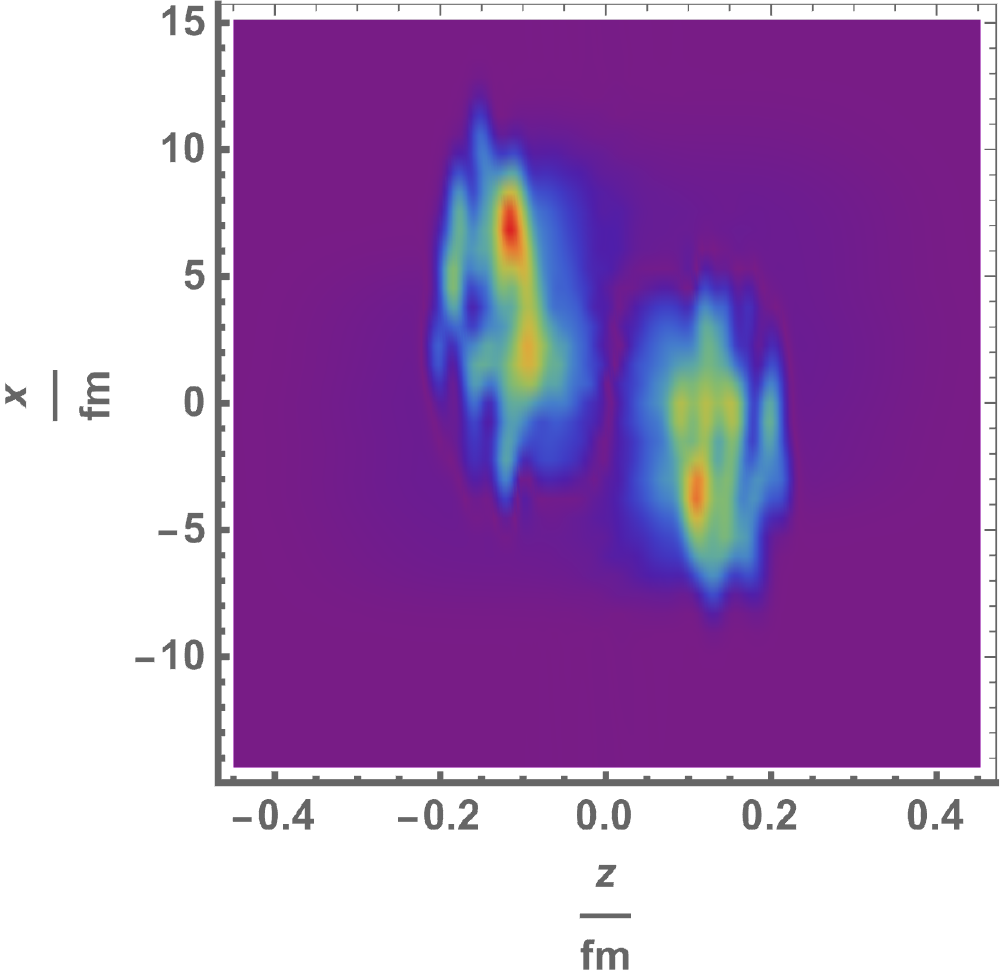}
\includegraphics[scale=0.6]{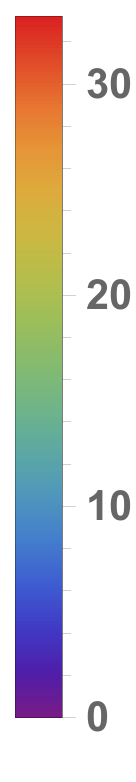}
\end{minipage}
\caption{The absolute value of the energy flux $|T^{0i}|$ in units of $[\text{GeV}^4]$ as a function of the longitudinal coordinate $z$ and the transverse coordinate $x$, at $y=0$ and at various times. From top left to bottom right, the $y=0$ slices depict the momentum density at $t= -0.144$ fm/$c$, $t= 0$ fm/$c$, $t= 0.068$ fm/$c$ and $t= 0.144$ fm/$c$.}
\label{momentum_slice}
\end{figure}

\begin{figure}
\begin{minipage}{0.5\textwidth}
\includegraphics[scale=0.6]{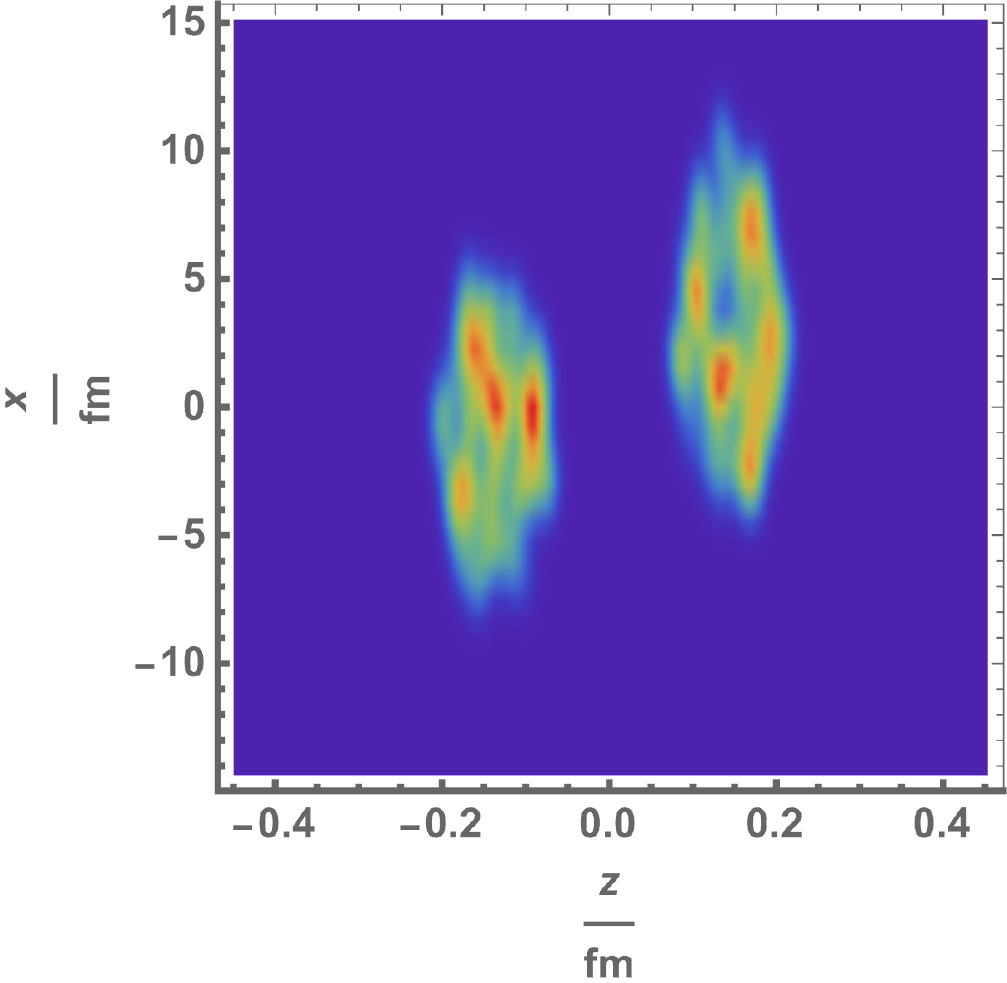}
\includegraphics[scale=0.6]{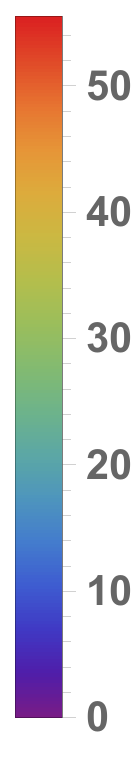}
\end{minipage}
\begin{minipage}{0.5\textwidth}
\includegraphics[scale=0.6]{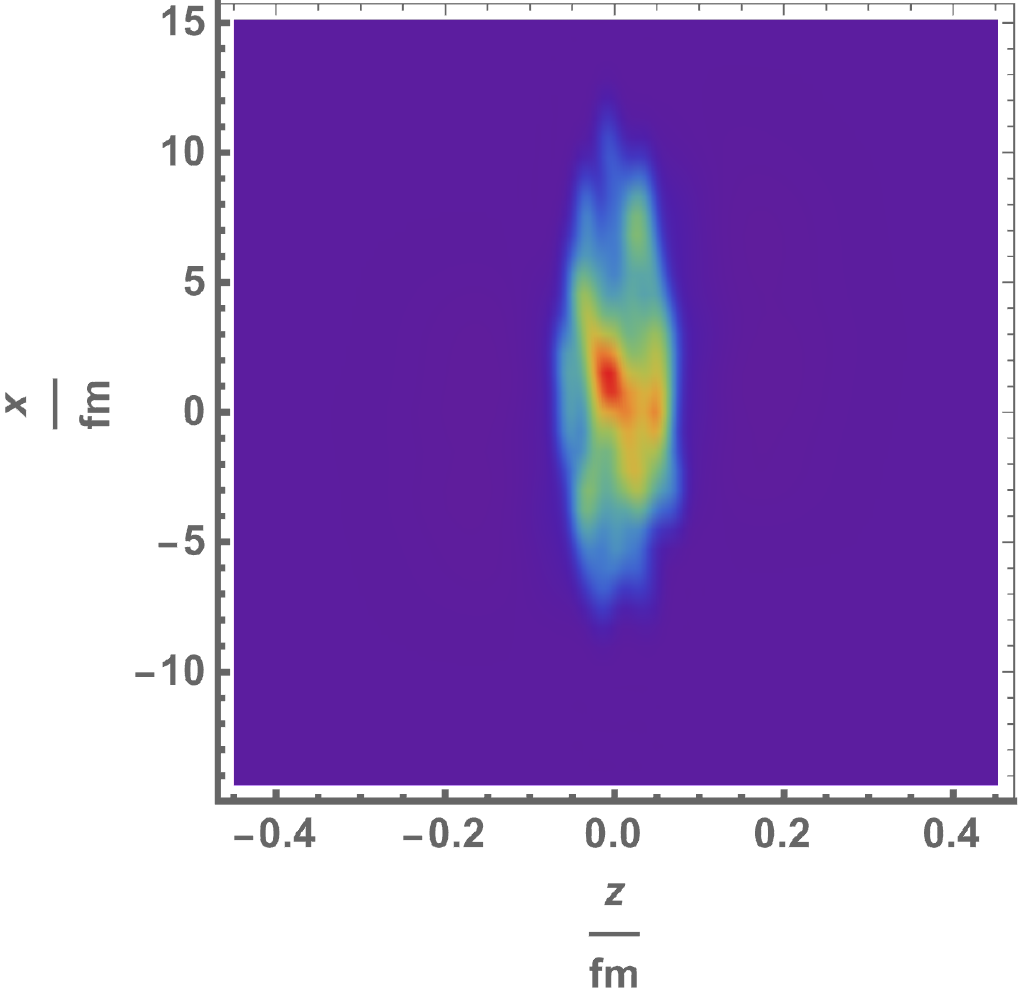}
\includegraphics[scale=0.6]{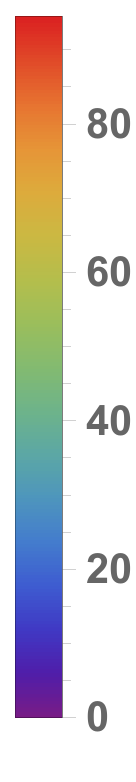}
\end{minipage}
\\\\\\
\begin{minipage}{0.5\textwidth}
\includegraphics[scale=0.6]{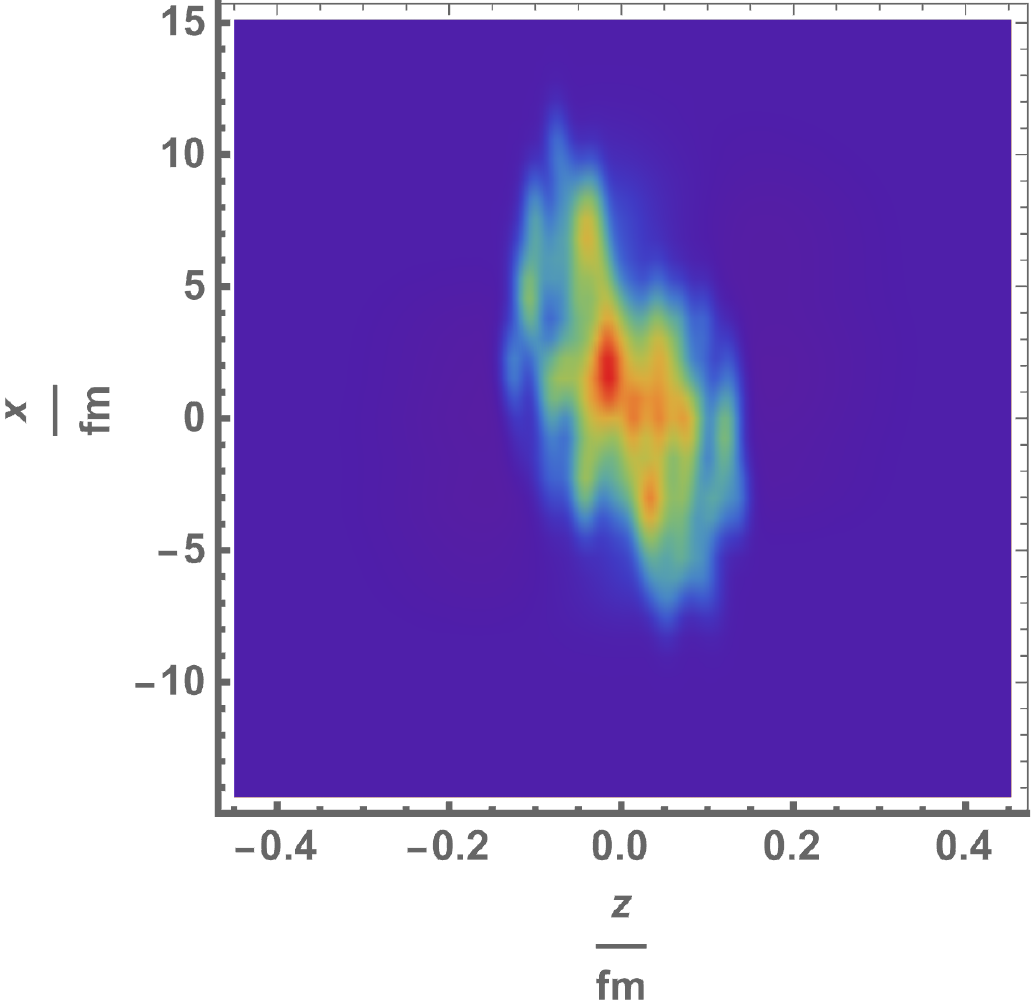}
\includegraphics[scale=0.6]{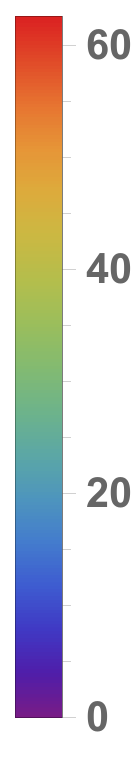}
\end{minipage}
\begin{minipage}{0.5\textwidth}
\includegraphics[scale=0.6]{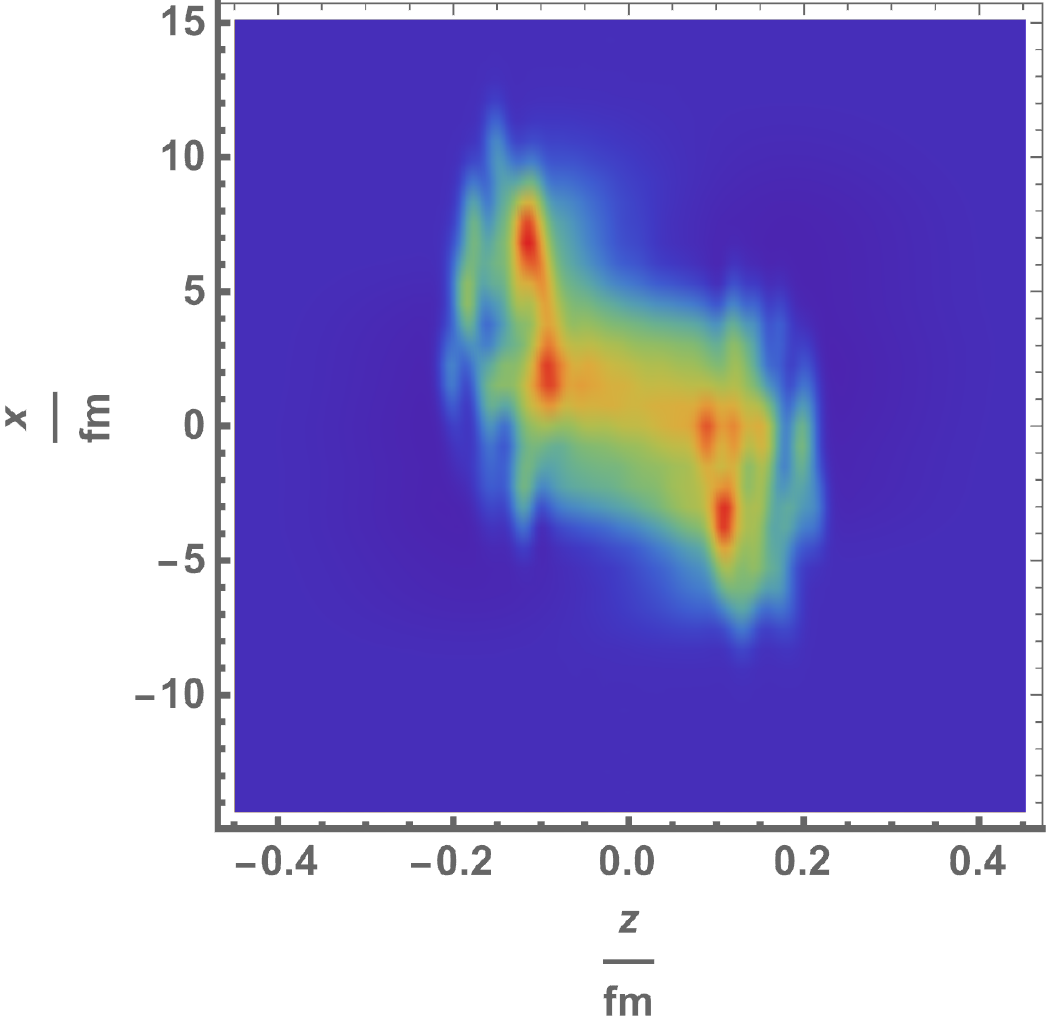}
\includegraphics[scale=0.6]{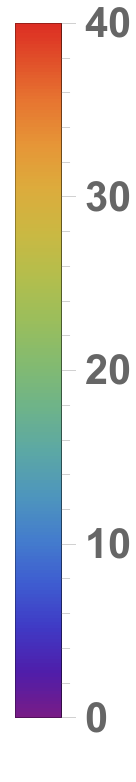}
\end{minipage}
\caption{The energy density $T^{00}$ in units of $[\text{GeV}^4]$ as a function of the longitudinal coordinate $z$ and the transverse coordinate $x$, at $y=0$ and at various times. From top left to bottom right, the $y=0$ slices are evaluated  at $t= -0.144$ fm/$c$, $t= 0$ fm/$c$, $t= 0.068$ fm/$c$ and $t= 0.144$ fm/$c$.}
\label{energy_slice}
\end{figure}

\begin{figure}
\includegraphics[scale=0.37]{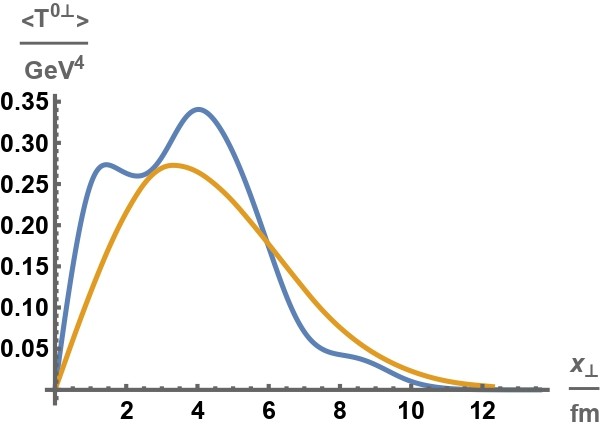}
\includegraphics[scale=0.37]{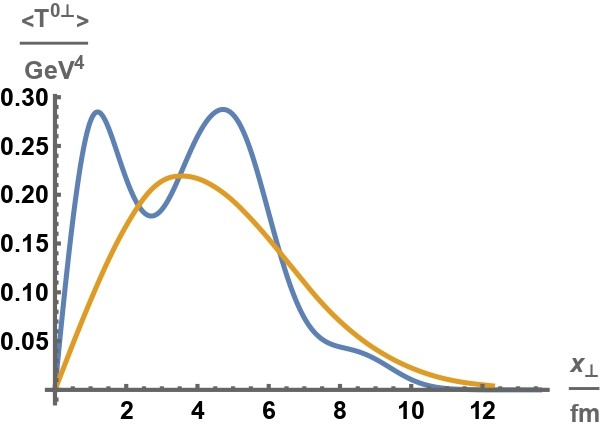}
\caption{The transverse plane angle-averaged transverse momentum density $\langle T^{0\perp} \rangle \equiv \langle T^{0 i} (\hat x^\bot)^i\rangle$ at rapidities $\xi=0$ (left) and $\xi=0.5$ (right), and at proper time $\tau =0.1$ fm/$c$. The blue curve corresponds to the results obtained in this work, starting from initial conditions (\ref{FG}) with left and right moving shocks given by (\ref{hpm}). \SW{ The yellow curves are obtained from extrapolating results in \cite{Che} using the fact that the difference between a first order in transverse derivative approximation and exact results for aspect ratios corresponding to those of Lorentz contracted projectiles at RHIC are small ($< 10 \%$) \cite{ 2206.01819}.} \iffalse For this we matched the amplitudes $\sqrt{\mu_+(\bold x_\bot=0)\,\mu_-(\bold x_\bot=0)}$, where $\mu_\pm(\bold x_\bot=0)^3$ is the  longitudinally integrated energy density at the central point of the right ($+$) and left ($-$) moving shocks. After this we rescale the transverse grid so that the transverse grid size in inverse units of $\sqrt{\mu_+(\bold x_\bot=0)\,\mu_-(\bold x_\bot=0)}$ in \cite{2206.01819} matches the size chosen in this work. Zeroth order in transverse derivative terms are not affected by changing the transverse length. First order quantities such as $\langle T^{0\perp} \rangle$ scale as  $\langle T^{0\perp} \rangle \rightarrow  a^{-1} \, \langle T^{0\perp} \rangle$ if we rescale the transverse length $L_\bot \rightarrow a L_\bot$. After these operations both the projectiles in  \cite{Che, 2206.01819} and in this work have a similar overlap region, a similar longitudinal width and by construction the same amplitude, making this comparison possible. Thus, the yellow curves in the plots above represent the prediction of \cite{Che} for the transverse flow during the early phase after heavy ion collisions, using realistic parameters for the transverse extent and the amplitude of the shocks. The blue curves are the updated results computed in this work, using a Woods-Saxon potential as probability distribution for the individual nuclei and taking into account  our particular realizations of the lumpy structure of the projectiles.\fi}
\label{transverse_flow}
\end{figure}

Using the gauge/gravity dictionary, we determine the boundary stress energy tensor from the near boundary expansion of the bulk metric. Figure \ref{contour_energy} depicts the energy density at \SW{$0.144$ fm/$c$ before the collision, at the time of the collision $t=0$ and at $0.068$ fm/$c$ and $0.144$ fm/$c$ after the collision, where we \SWn{chose} $t=0$ as the time when the center of masses of the projectiles are located at the same longitudinal coordinate. At the time of the collision\SWn{, $t=0$, } the maximum energy density has reached $160\%$ of the peak energy density of an initial projectile. Due to the granularity of the initial data, the overall maximum energy density is not reached exactly at  $t=0$, but at $t= 0.034$ fm/$c$ after the collision and measures $170\%$ of the initial peak energy density. At $t=0$, similar to what is observed during planar collisions and in \cite{Che}, the energy density profile matches to a good accuracy (with an error of approximately $ 0.15\%$) \SWn{the} superposition of the two initial shocks. The second row of figures in Fig.~\ref{contour_energy} shows the energy density at times $0.068$ fm/$c$ and $0.144$ fm/$c$ after the collision.
The maximum of the  energy density on those times slices has decreased to $107\%$ and \SWn{$50\%$} of the initial peak energy density, respectively. \SW{Towards the endpoint of our time integration at} time $t=0.144$ fm/$c$ the energy density averaged over the central region ($|x_\bot|<2.5$ fm) falls off with the approximate rate $\propto t^{-0.9}$, the same rate  as observed during planar collisions. }
In Fig \ref{momentum_slice} and Fig.~\ref{energy_slice} we show the momentum density and the energy density \SW{at vanishing $y$ coordinate, where $\hat y$ is the transverse plane unit vector orthogonal to the impact parameter vector,
on the same time slices as depicted in Fig.~\ref{contour_energy}. } Figure \ref{transverse_flow} shows the lab-frame angle-averaged transverse energy flux, \SW{ $\langle T^{0\bot} \rangle \equiv \langle T^{0i} (\hat x^\bot)^i\rangle$}, as a function of the transverse radius $x_\bot\equiv \sqrt{x^2+y^2}$. \SW{We also compare our results to previous ones for the transverse flow, where the granular structure of the projectiles had not been taken into account \cite{Che}. For this we matched the amplitudes $\sqrt{\mu_+(\bold x_\bot{=}0)\,\mu_-(\bold x_\bot{=}0)}$, where $\mu_\pm(\bold x_\bot{=}0)^3$ is the  longitudinally integrated energy density at the central point of the right ($+$) and left ($-$) moving shocks. After this we rescale the transverse grid so that the transverse grid size in inverse units of $\sqrt{\mu_+(\bold x_\bot=0)\,\mu_-(\bold x_\bot=0)}$ in \cite{2206.01819} matches the size chosen in this work.  Zeroth order in transverse derivative terms are not affected by changing the transverse length. First order quantities such as $\langle T^{0\perp} \rangle$ scale as  $\langle T^{0\perp} \rangle \rightarrow  a^{-1} \, \langle T^{0\perp} \rangle$ if we rescale the transverse length $L_\bot \rightarrow a L_\bot$. Note that starting from a solution to the Einstein equations and rescaling both the amplitude and  the transverse size, without changing \SWn{the} longitudinal size, does not in general generate \SWn{ a  valid  solution} of the Einstein equations. Therefore, a priori it was not clear whether the results obtained in  \cite{Che} can be used to approximate collisions with realistic aspect ratios of the colliding projectiles, corresponding to Lorentz contractions at RHIC, without showing that the disagreement between the first order in derivative approximation and exact results is small \cite{2206.01819}. 
 After these operations both the projectiles in  \cite{Che, 2206.01819} and in this work have a similar overlap region, a similar longitudinal width and by construction the same amplitude, making this comparison possible. \SWn{The} yellow curves in Fig \ref{transverse_flow} represent the prediction of \cite{Che} for the transverse flow during the early phase after heavy ion collisions, using realistic parameters for the transverse extent and the amplitude of the shocks. The blue curves are the updated results computed in this work, using a Woods-Saxon potential as probability distribution for the individual nuclei and taking into account  our particular realizations of the lumpy structure of the projectiles.  While the maximum value and the compact support of the averaged transverse energy flux of the results in \cite{Che,2206.01819} and the results presented in this work (that take into account the granular structure of the projectiles) are similar, the shape of $\langle T^{0\bot} \rangle$ as a function of $x_\bot$ is noticeably affected by starting from lumpy instead of smooth initial conditions, despite the angle average. }

\FloatBarrier
\subsection{Hydrodynamization}
Comparing the stress energy tensor after the collision with its hydrodynamic approximation, where the constitutive relations are truncated after the first order in derivatives, allows one to quantify whether a hydrodynamic description of the dynamics is useful. At each order in the transverse derivative expansion we compute the fluid velocity from the eigenvalue equation
\begin{equation}
T^\mu\,_\nu \,u^\nu = -\varepsilon \, u^\mu \,,
\label{Eigenvalue_fluid_velocity}
\end{equation}
where the eigenvalue $\varepsilon$ is
 the proper energy density.
The hydrodynamic approximation 
\begin{equation}
    \widehat T^{\mu\nu}_{\rm hydro}
    = p \, g^{\mu\nu} + (\varepsilon{+}p) \, u^\mu u^\nu + \Pi^{\mu \nu} \,,
    \label{hydro_approx}
\end{equation}
with the viscous stress $\Pi$ given by
\begin{equation}
    \Pi_{\mu\nu}
    =
    -2\,\eta \,
    \left[
    \partial_{(\mu}u_{\nu)}
    + u_{(\mu} u^\rho \partial_\rho u_{\nu)}
    - \tfrac 13 \,  \partial_\alpha u^\alpha
    (\eta_{\mu\nu} + u_\mu u_\nu)
    \right]
    + \mathcal{O}(\partial^2) \,,
\end{equation}
is also expanded up to first order in transverse derivatives. Here $p$ is the pressure  and $\eta$ the shear viscosity. 
\begin{figure}
\includegraphics[scale=0.7]{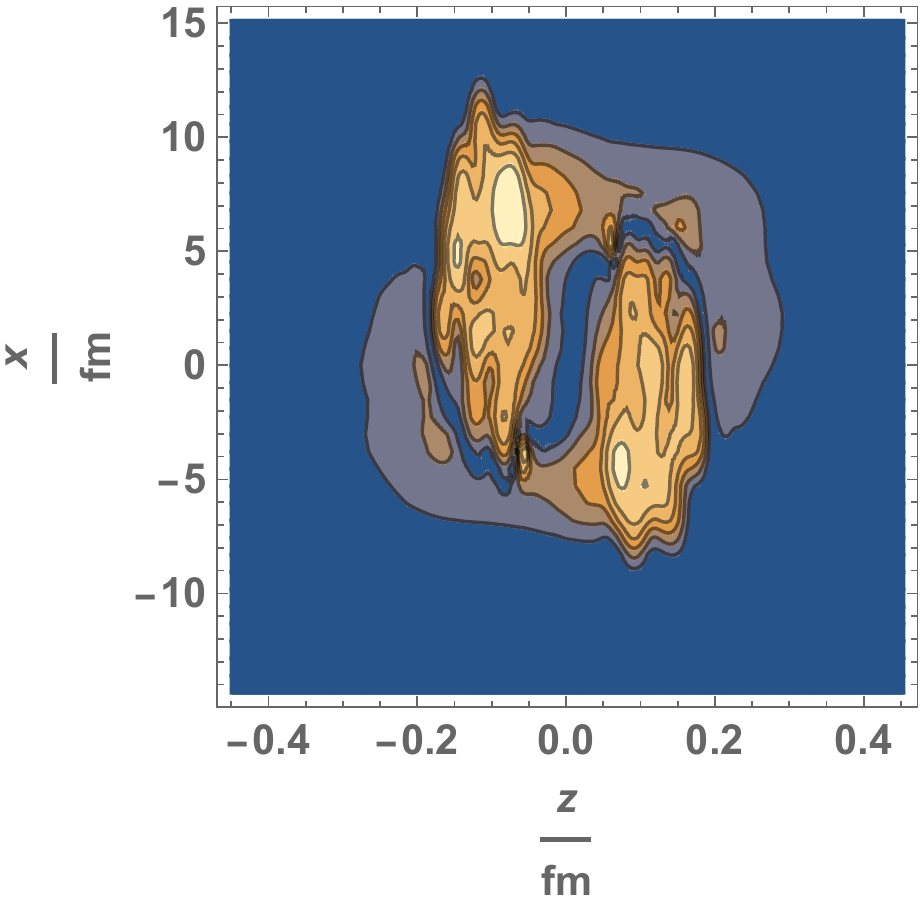}
\includegraphics[scale=0.5]{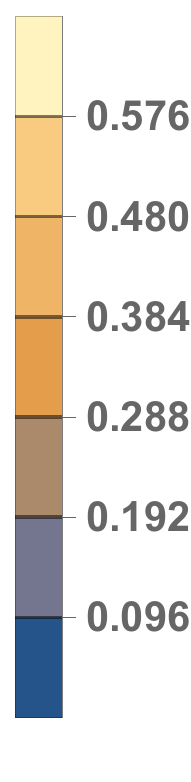}
\includegraphics[scale=0.7]{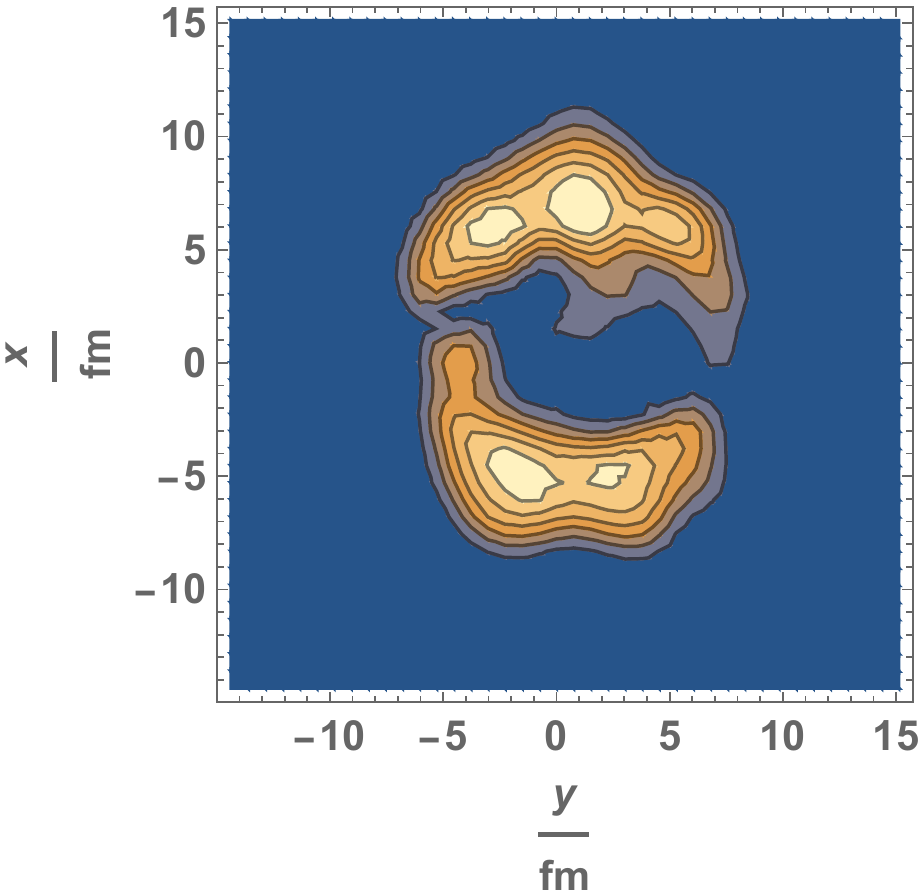}
\includegraphics[scale=0.5]{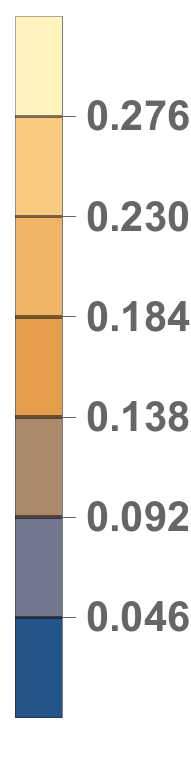}
\caption{The absolute value of the fluid three velocity $|\bold u /u^0|$  at time $t = 0.1$ fm/$c$. The left plot shows  $|\bold u /u^0|$   at a $y=0$ slice, the right plot shows it at a $z=0$ slice.}
\label{fluid_velo}
\end{figure}

We show slices of the fluid velocity three vector's absolute value $|\bold u /u^0|$  at time $t = 0.1$ fm/$c$ in Fig.~\ref{fluid_velo}. Next we compute the  residual
\begin{equation}
\Delta = \frac{3}{\varepsilon} \sqrt{\Delta T^{\mu \nu} \Delta T_{\mu \nu}}
\label{residual}
\end{equation}
with $\Delta T^{\mu \nu} = T^{\mu \nu}- \widehat T^{\mu\nu}_{\rm hydro}$. Following earlier work \cite{che3, Che, Chesler:2015fpa}, $\Delta< 0.15$ is regarded as the onset of approximate validity of hydrodynamics. As shown in \cite{2206.01819}, first order corrections to the residual $\Delta$ are negligible. However, explicitly computing first order in transverse derivative corrections of the fluid velocity from Eq.~(\ref{Eigenvalue_fluid_velocity}) is necessary for determining the vorticity (up to first order in transverse gradients), which is discussed in the next section.

\begin{figure}
\includegraphics[scale=0.65]{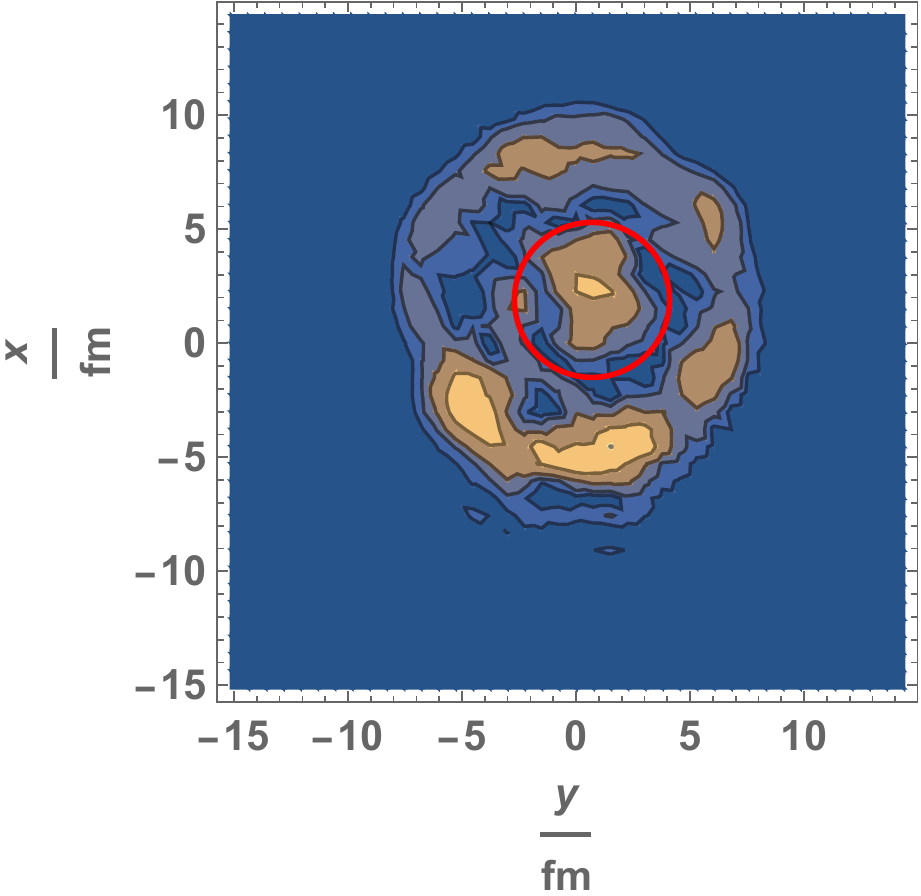}
\includegraphics[scale=0.5]{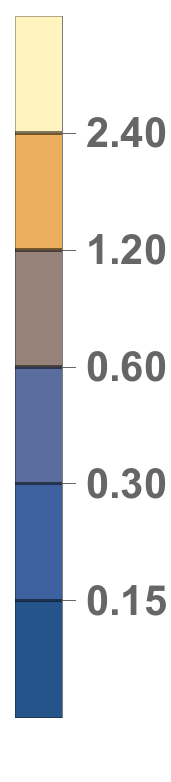}
\includegraphics[scale=0.65]{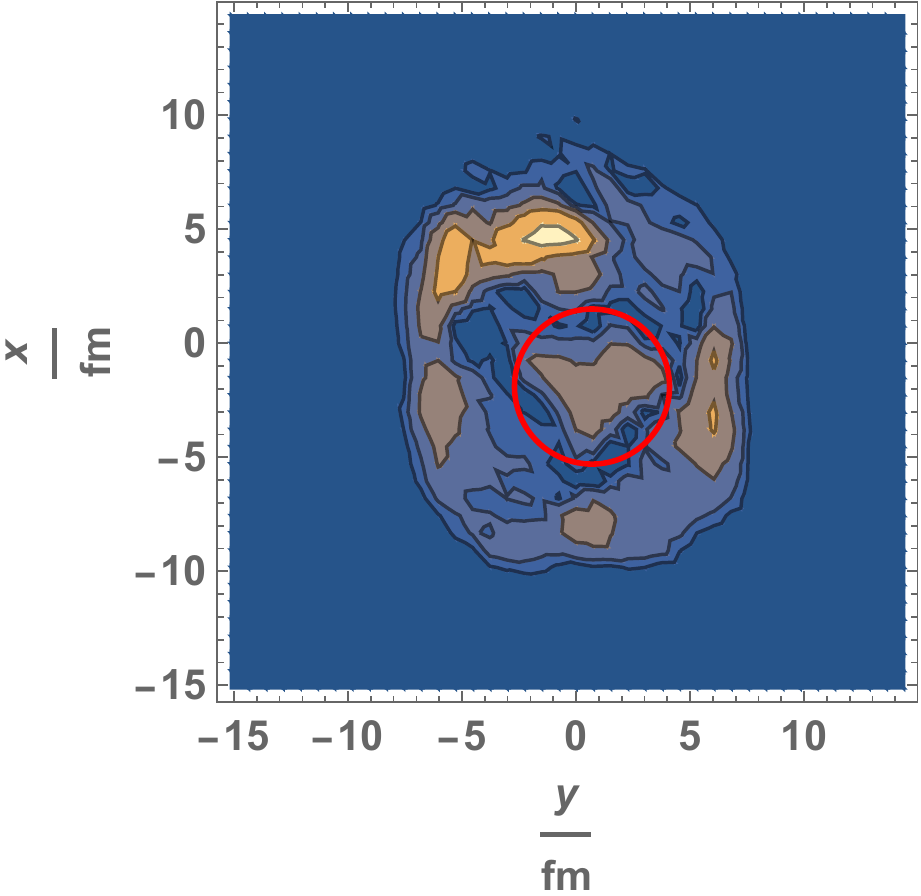}
\includegraphics[scale=0.5]{delta_2_s.pdf}
\begin{center}
\includegraphics[scale=0.65]{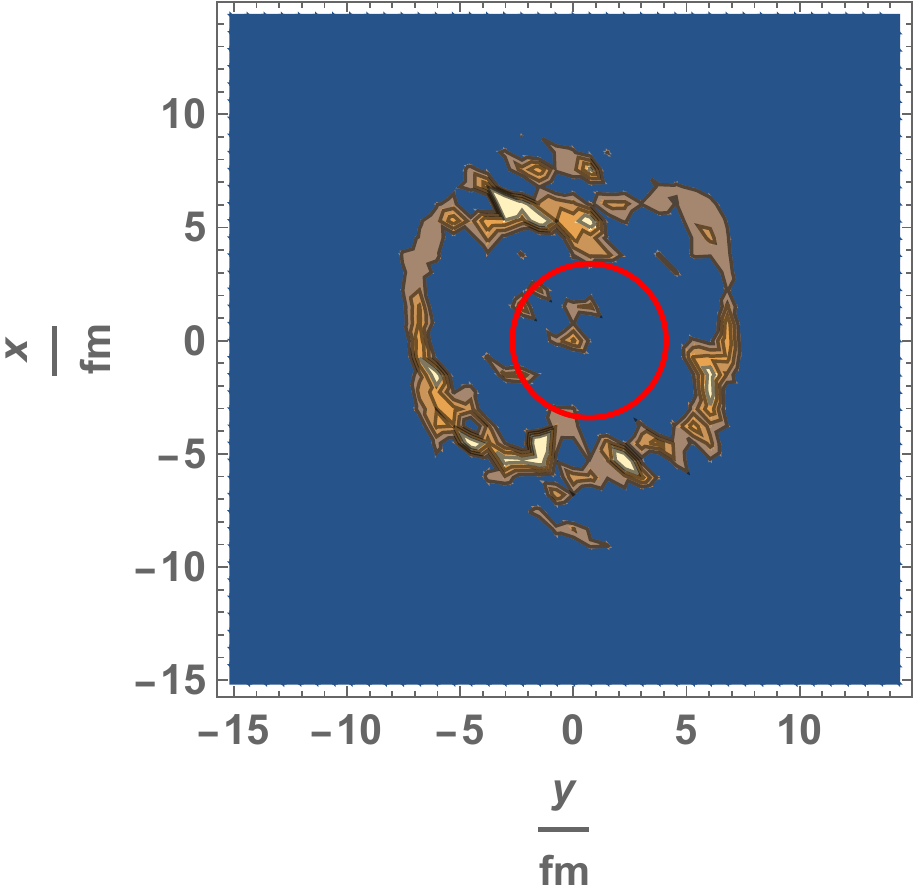}
\includegraphics[scale=0.5]{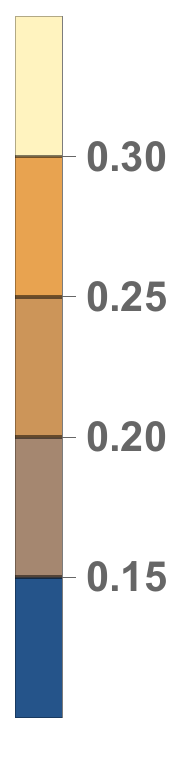}
\end{center}
\caption{The residual $\Delta$   at proper time \SW{$\tau = 0.106$ fm$/c$}. For rapidity  $\xi=-0.5$ (left) and  $\xi=0.5$ (right) in the first row and vanishing rapidity for the plot in  the second row. \iffalse  While most of  the low rapidity, central region can be described by hydrodynamics at  $\tau = 0.1$ fm$/c$, only a small subset of the plasma at mid-rapidity has hydrodynamized at this proper time. In order to provide initial data for hydro evolutions on a full initial hypersurface, one would have to evolve the geometry substantially longer, which goes beyond the scope of this work. We can, however, identify a rapidity dependent ring shaped subregion $\mathcal{R}$ in the transverse plane given in units of [fm] by $\bold x_\bot = \{0.7 + 3.4 \, \sin(\phi), -3.8 \, \xi + \,\cos(\phi)\}$  with $\xi \in [0,0.5]$, $\phi \in [0, 2 \pi]$, where the plasma can be described by hydrodynamics already at proper time $\tau=0.1 $ fm/$c$.\fi  We display the region $\mathcal{R}$ described in (\ref{R}) by a red circle in the plots above. }
\label{Delta}
\end{figure}

\begin{figure}
\begin{center}
\hspace{0.35 cm}\includegraphics[scale=0.65]{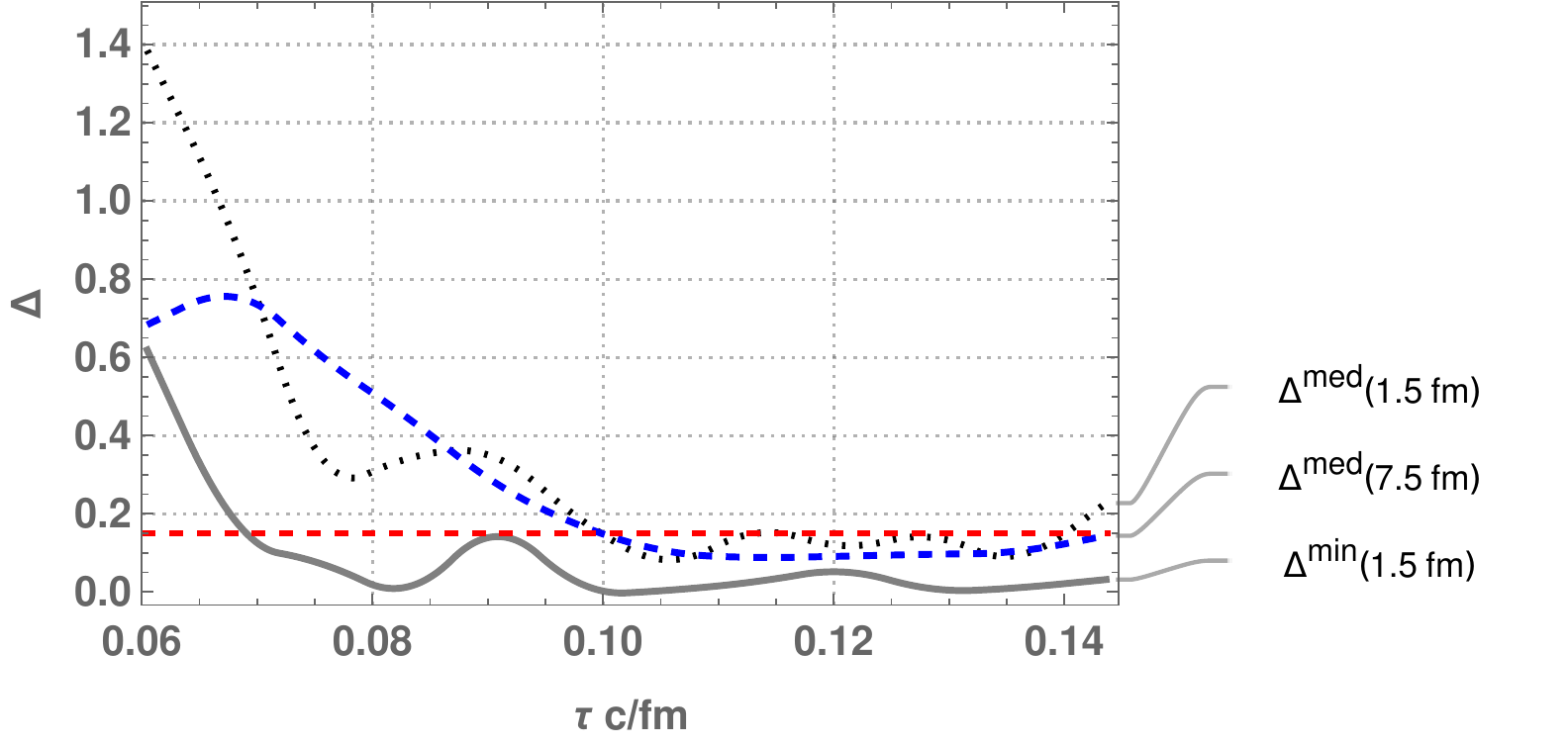}\\
\includegraphics[scale=0.65]{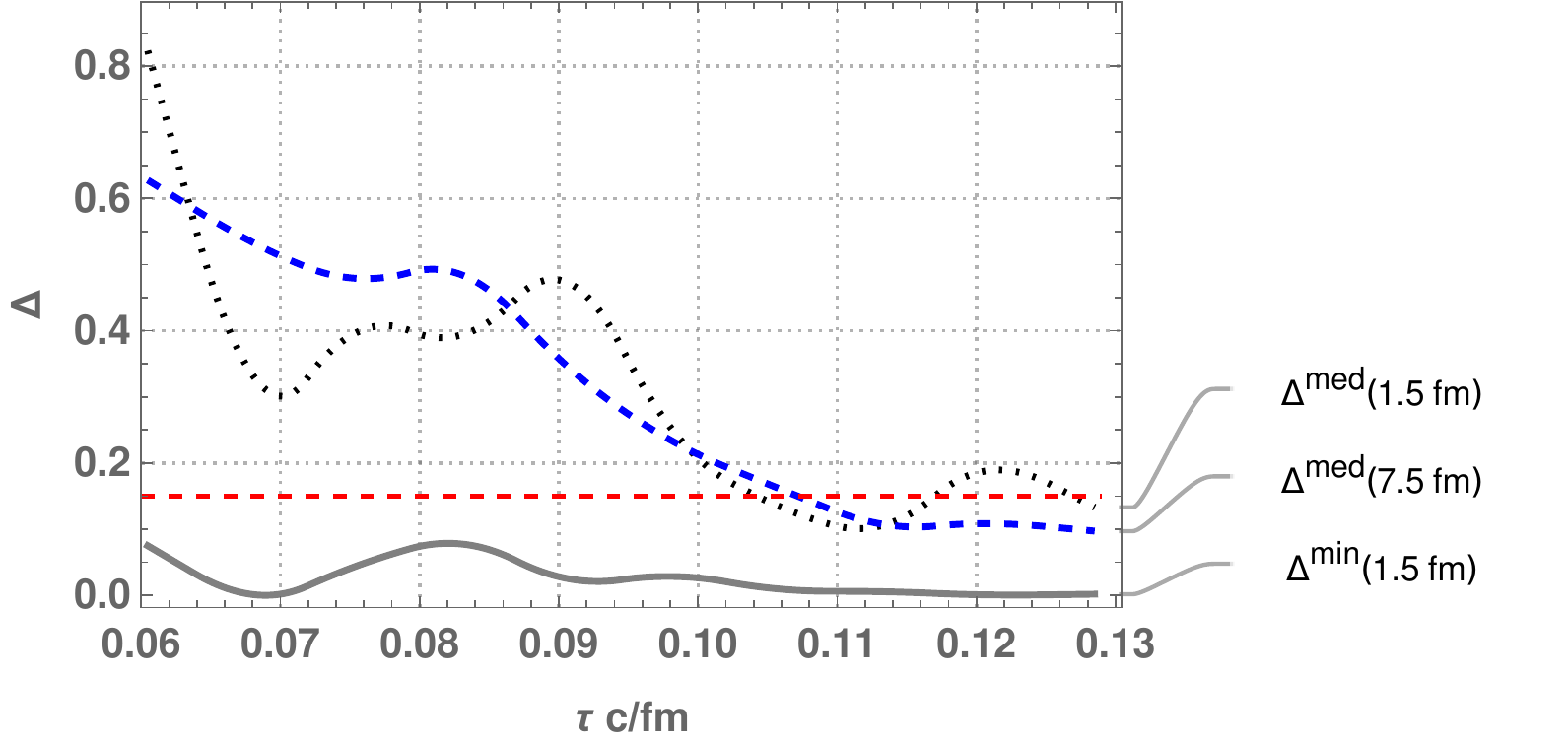}
\end{center}
\caption{\SW{In the top plot we show the median hydro residual at rapidity $\xi=0$. The red dashed line corresponds to $\Delta =0.15$. The black dotted line shows the residual averaged over the central region $|x_\bot|<7.5$ fm, whereas the blue dashed curve shows the same for $|x_\bot|<1.5$ fm. The gray line shows the minimal $\Delta$ in the region $|x_\bot|<1.5$ fm. The plot below depicts the analogous functions at rapidity $\xi=0.25$.}}
\label{Delta2}
\end{figure}

\begin{figure}
\begin{center}
\includegraphics[scale=0.65]{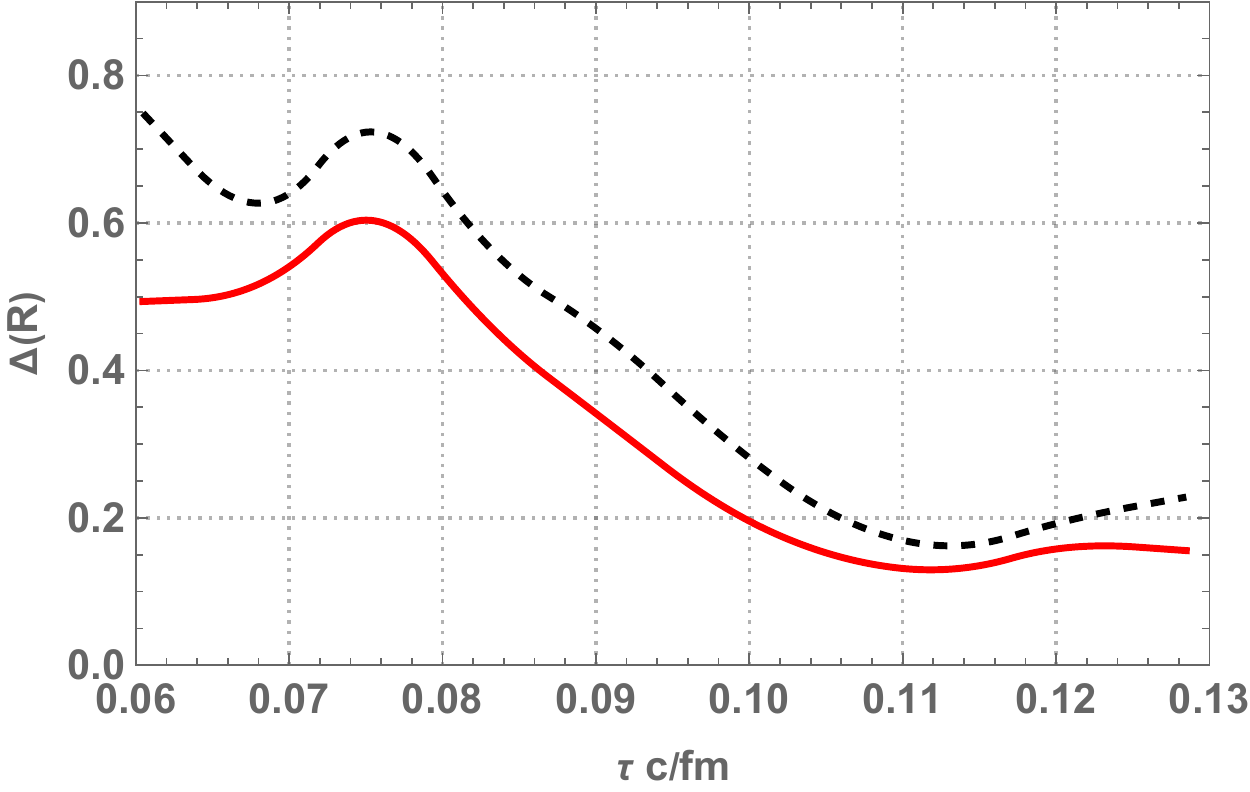}
\end{center}
\caption{\SW{The hydro residual $\Delta(\mathcal{R})$ averaged over the region $\mathcal{R}$ as a function of proper time $\tau$ (black, dashed curve) and the  median $\Delta$ in the region $\mathcal{R}$ (red solid line).} }
\label{Delta3}
\end{figure}

\begin{figure}
\begin{center}\hspace{0.2cm}
\includegraphics[scale=0.51]{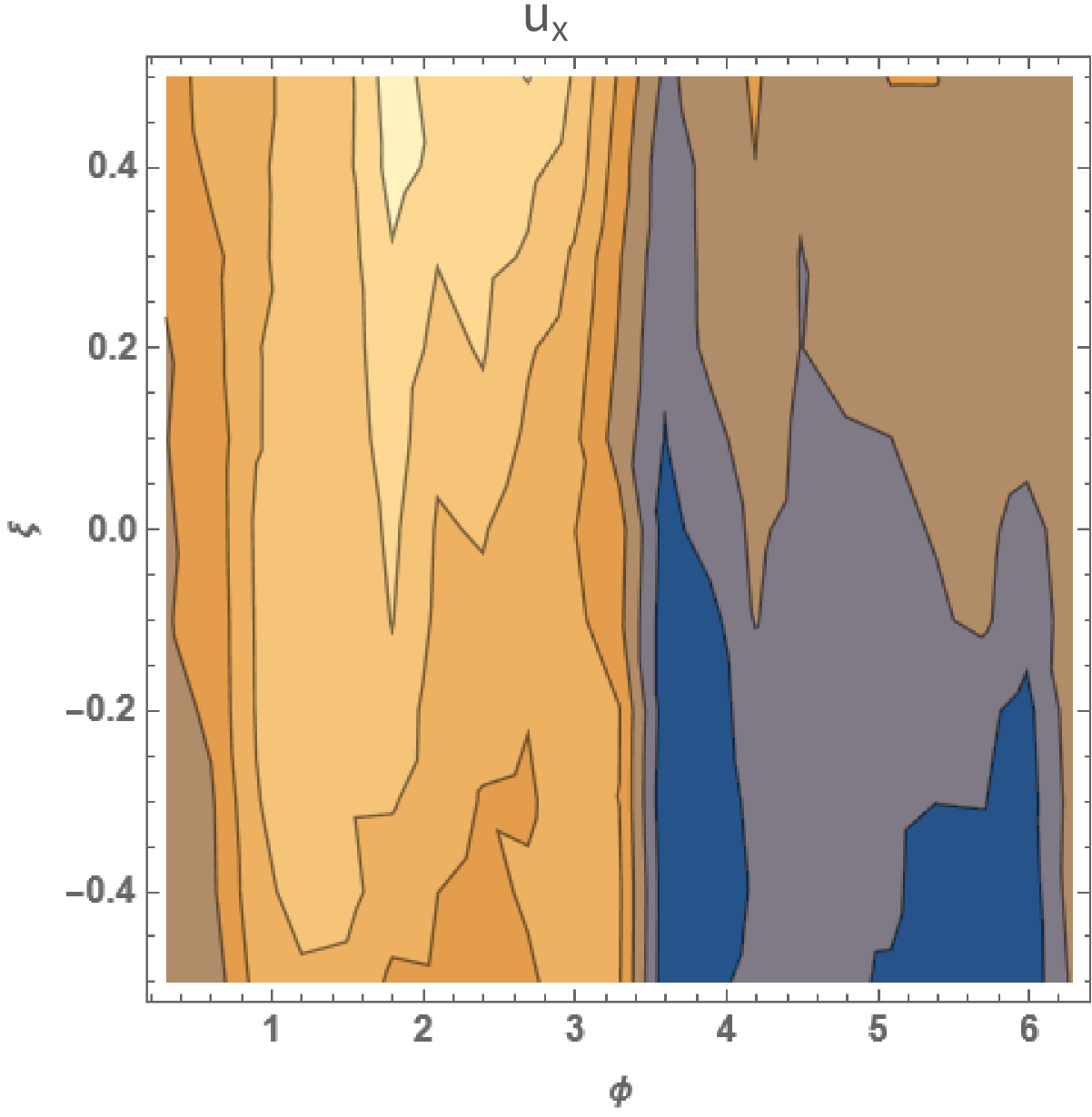}
\includegraphics[scale=0.5]{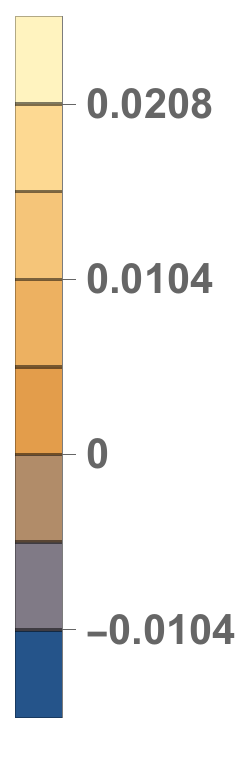}
\includegraphics[scale=0.51]{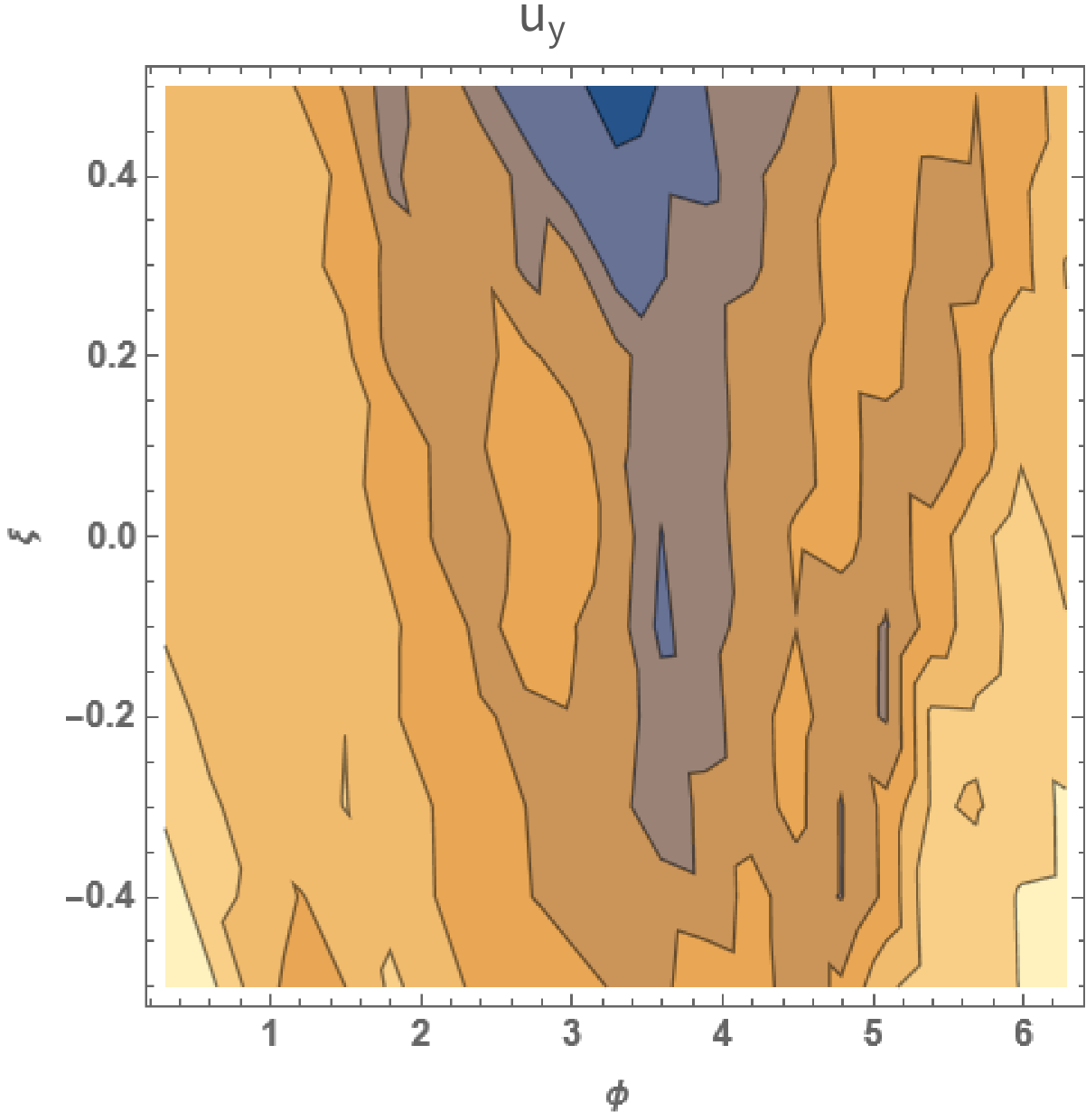}
\includegraphics[scale=0.5]{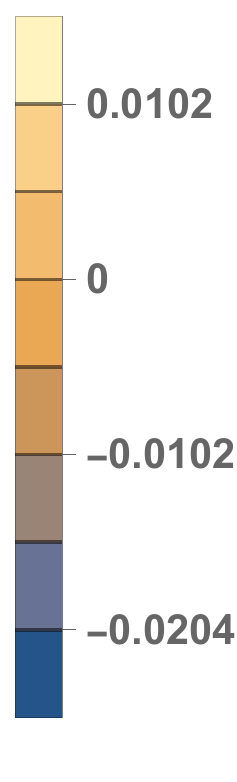}
\includegraphics[scale=0.51]{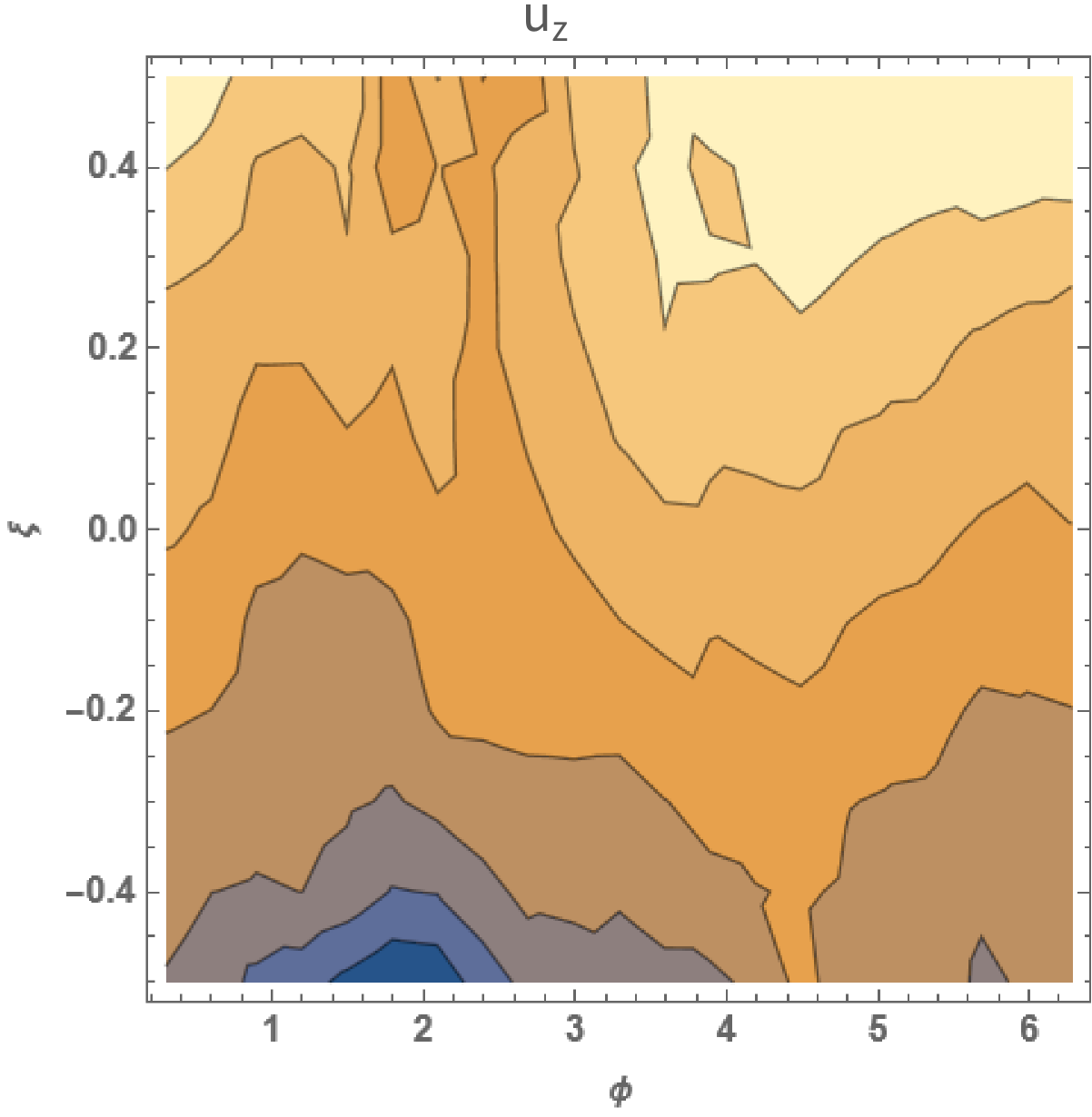}
\includegraphics[scale=0.5]{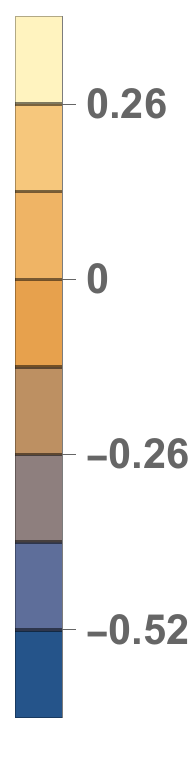}
\includegraphics[scale=0.51]{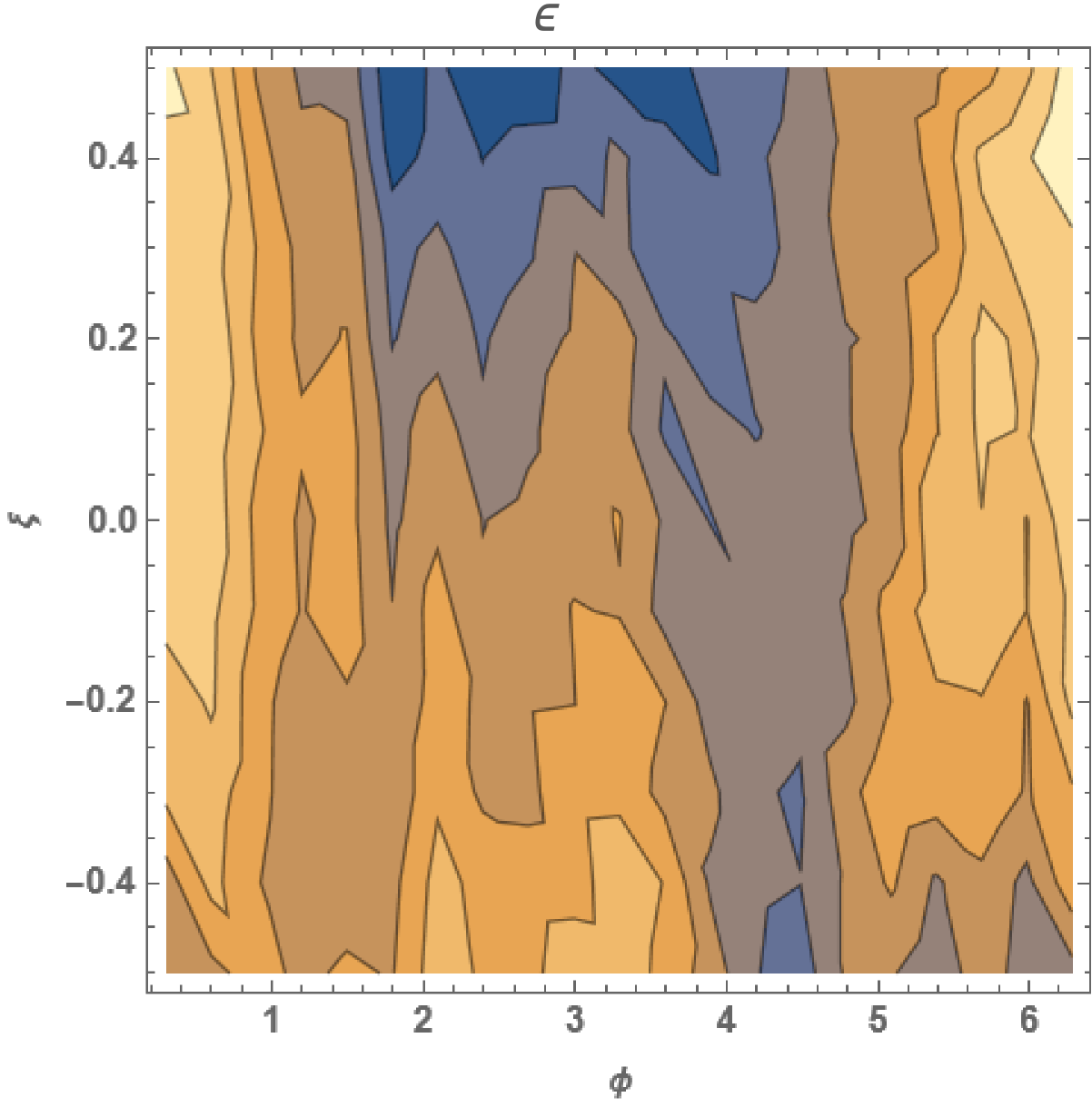}
\includegraphics[scale=0.5]{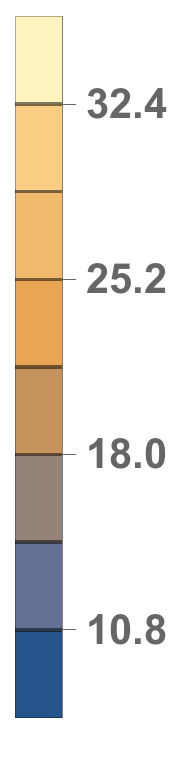}
\caption{The fluid velocity and the proper energy density in the hydro-subregion $\mathcal{R}$ defined in (\ref{R}) as a function of rapidity $\xi$ and $\phi$ given in (\ref{R}). The first row corresponds to the transverse fluid velocity with $u_x$ corresponding the left and $u_y$ to the right plot. The second row shows the longitudinal fluid velocity $u_z$ on the left and the proper energy density, which is given in units of [GeV$^4$], on the right. }
\label{R_results}
\end{center}
\end{figure}

We show the results for $\Delta$ in Fig.~\ref{Delta}. As can be seen there, most of  the low rapidity  ($\xi \approx 0$) central region can be described by hydrodynamics at  $\tau = 0.1$ fm$/c$, but only a small subset of the plasma at mid-rapidity ($|\xi| \approx 0.5$) has hydrodynamized at this proper time. In order to provide initial data for hydro evolutions on a full initial hypersurface, one would have to evolve the geometry substantially longer, which goes beyond the scope of this work. \SWn{At time $t\approx 0.1$ fm$/c$ after the collision, the majority of the plasma around  the central point $\bold x_\bot=0$, $z=0$ is hydrodynamized. We depict this behavior in Fig.~\ref{Delta2}, where  we show the median of the hydro residual  $\Delta$  in the central regions $|x_\bot|<1.5$ fm and $|x_\bot|<7.5$ fm both at rapidity $\xi=0$ and at rapidity $\xi=0.25$ as a function of proper time. As shown there at proper time $\tau \approx 0.1$ fm$/c$, the majority of the low rapidity plasma in the central region has hydrodynamized, while even at vanishing rapidity individual transverse pixels can still be far from the hydrodynamic approximation, as shown in Fig.~\ref{Delta}.  Nonetheless, we can attempt to identify the early proper time part of the hydrodynamization surface:} In the immediate neighborhood of the tube \SW{or thin pipe shaped} subregion $\mathcal{R}$ defined at constant proper time $\tau=0.106 $ fm/$c$ via
\begin{equation}
\bold x_\bot = \{ -3.8 \, \xi + 3.4\,\cos(\phi), 0.7 + 3.4 \, \sin(\phi)\},
\label{R}
\end{equation}
 for $\phi \in [0,2\pi]$ and $|\xi|<0.5$, \SWn{the median hydro residual is already below the threshold $\Delta < 0.15$.}
 \SWn{The red circle in  Fig.~\ref{Delta} shows slices of this region $\mathcal{R}$ at $\tau =1.06$.}  \SW{It should be noted that, due to the strong inhomogeneity of the hydro residual $\Delta$, statements about the exact hydrodynamization time can only be made locally or by averaging. They generally depend on the specific spatial region that is considered. The origin of this inhomogeneity is well understood: As observed in \cite{Chesler:2015fpa,wae3}, the hydrodynamization proper time of the plasma located at a transverse pixel is proportional to the inverse geometric mean of the longitudinally integrated energy densities of the two initial projectiles evaluated at the transverse coordinate of this pixel. Thus the hydro residual $\Delta$ reflects the strong transverse fluctuations of the initial data. \SWn{ In Fig.~\ref{Delta3} we depict  both the average and the median  hydro residual $\Delta$ in the region $R$ as a function of  proper time.} At proper time  $\tau = 1.06$ fm$/c$ the averaged hydro residual has dropped to $0.19$, whereas the median is at $0.14$.}
\SW{In \LY{summary}, the median of the hydro residual $\Delta$ in the central region (shown in Fig.~\ref{Delta2}) \LY{drops} below $0.15$ at time $t= 0.1$ fm$/c$ which, in units of the longitudinally integrated energy density of the central region $\mu^3$, corresponds to $t=1.4/\mu$, and is therefore close to the hydrodynamization time of the central region found in \cite{che3} without  granular initial initial data, \LY{which was $t \approx 1.25/\mu$}.
However, the hydrodynamization time of individual pixels varies drastically due to the influence of the granular structure on the local energy density scale,
\LY{as is clearly evident from the variation in the hydro residual
$\Delta$ shown in Fig.~\ref{Delta}}.
This is in line with expectation of earlier works \cite{mue} which predicted a substantial delay (by roughly a factor of 2) of the hydrodynamization of the full system due to the granular structure.}
  We show the fluid velocity and the proper energy density in the region $\mathcal{R}$ as function of the angle  $\phi$ and the rapidity $\xi$ in Fig.~\ref{R_results}. One sees that the transverse velocity in this region is quite modest, $|\vec u_\bot| \lesssim 0.02$ while the longitudinal velocity component is substantial, $|u_z| \sim 0.25$ at rapidity $\xi = \pm 0.5$.

\subsection{Vorticity}

\begin{figure}
\includegraphics[scale=0.65]{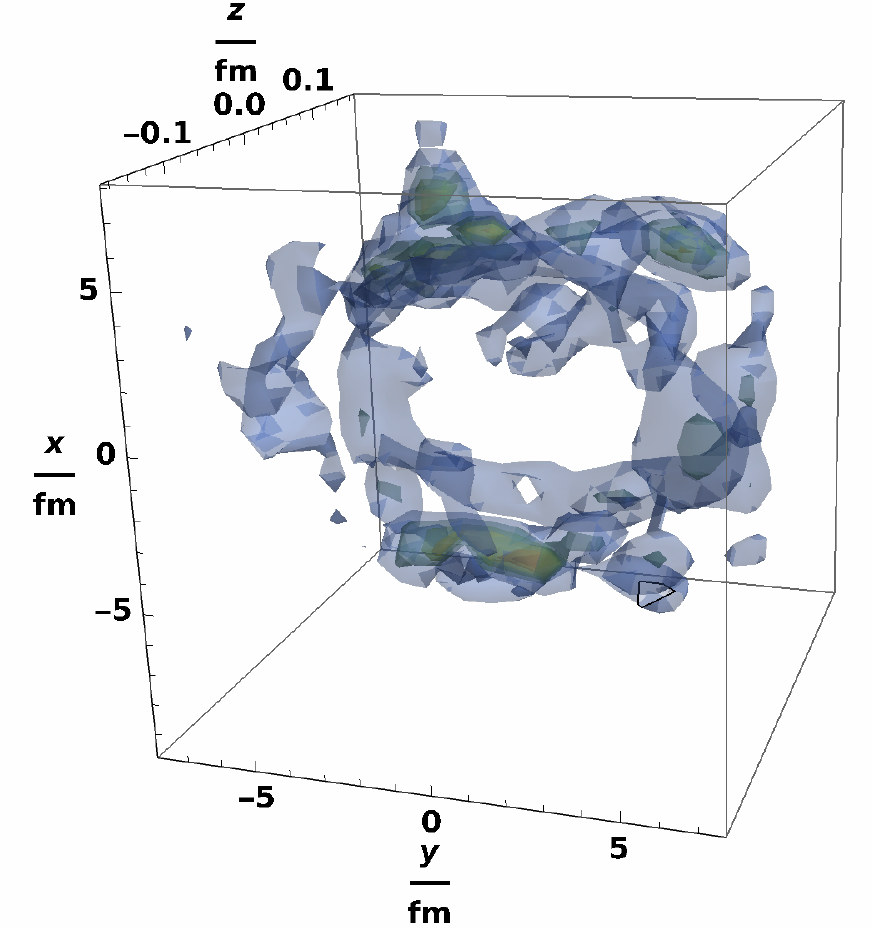}
\includegraphics[scale=0.65]{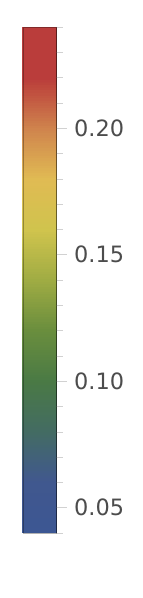}
\hspace{0.5cm}
\includegraphics[scale=0.7]{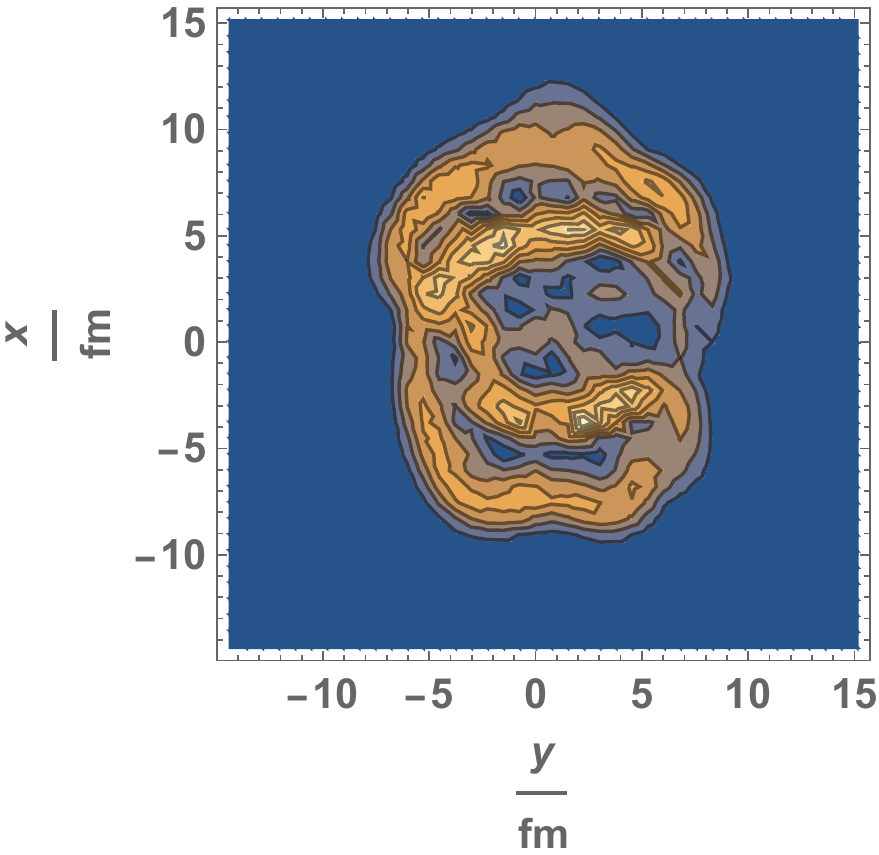}
\includegraphics[scale=0.55]{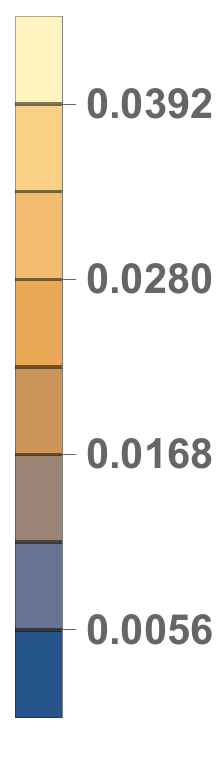}
\caption{On the left: The absolute value of the vorticity three vector  $\vec \omega$ at $t = 0.1$ fm$/c$, the spatial coordinates on the axes are given in units of [fm]. On the right we show the vorticity at $t = 0.1$ fm$/c$ at vanishing rapidity. The results are given in units of $[\text{GeV}]$. Most of the initial (geometric) angular momentum is deposited far away from the central region where the hydrodynamized quark gluon plasma is located.}
\label{fig_vort}
\end{figure}

\begin{figure}[htp]
\begin{center}
\vspace*{-0.1cm}
\includegraphics[scale=0.65]{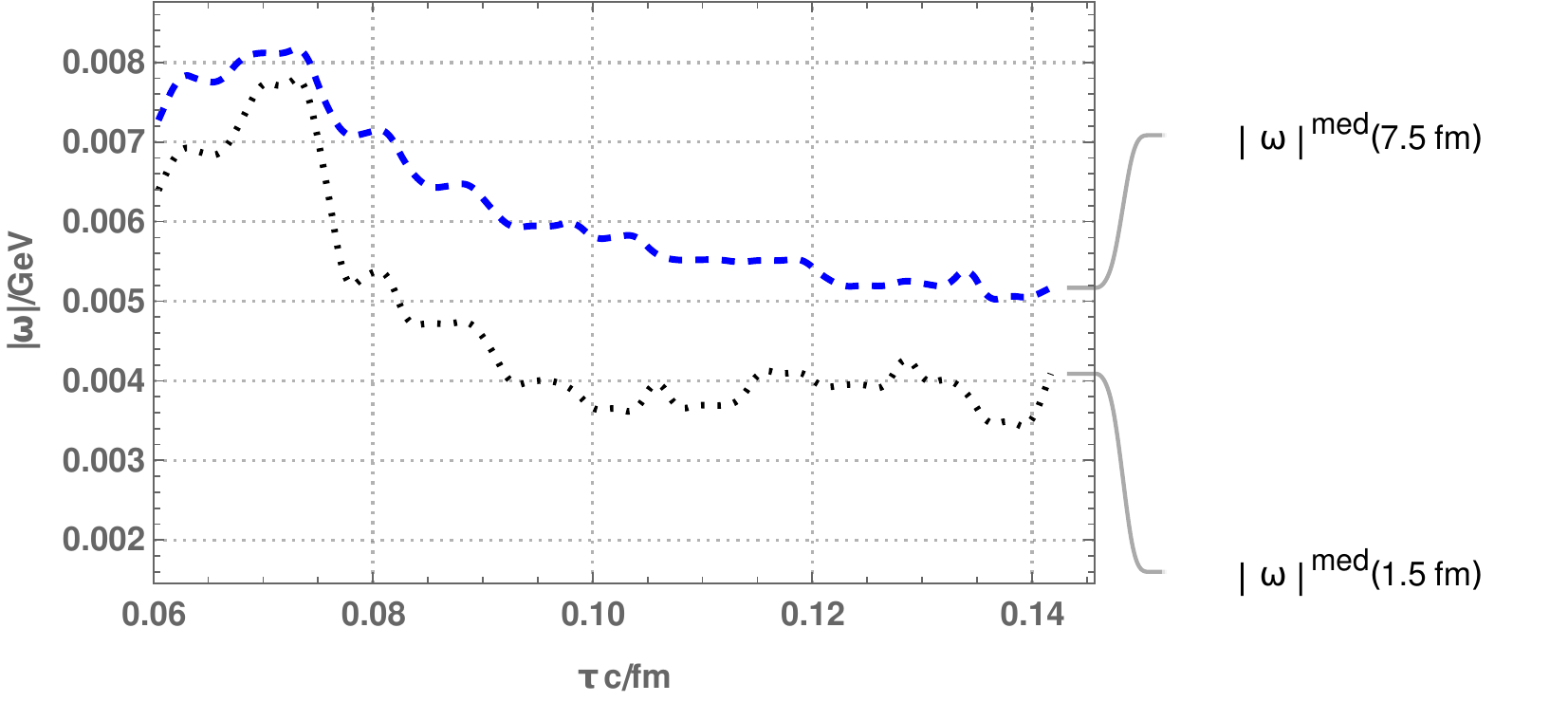}
\\
\hspace{0.25cm}
\includegraphics[scale=0.65]{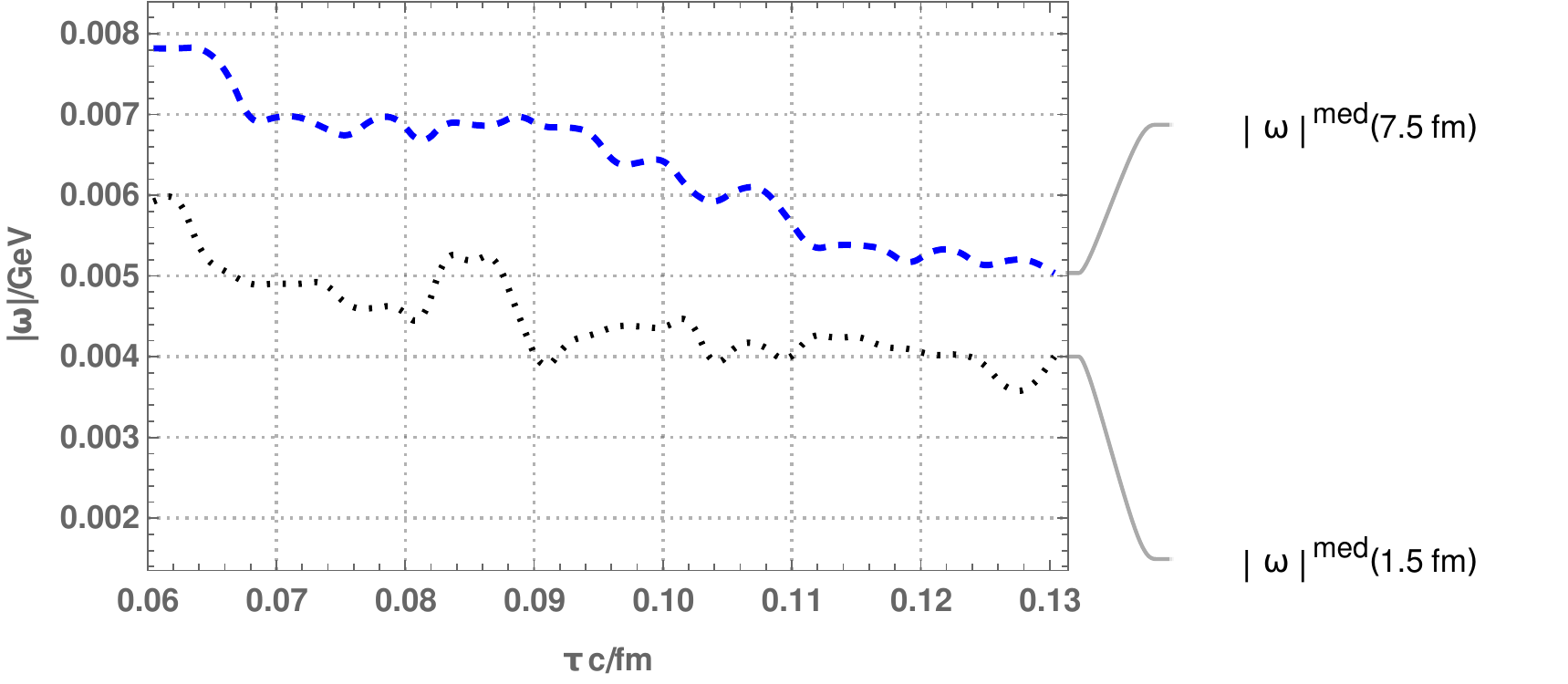}
\end{center}
\vspace{-0.6cm}
\caption{\SW{In the top plot we show the median absolute value of the three vector  vorticity $|\vec{\omega}|$, given by the spatial components of (\ref{vort}) at rapidity $\xi=0$. The black dotted line shows the vorticity averaged over the central region $|x_\bot|<7.5$ fm, whereas the blue dashed curve shows the same for $|x_\bot|<1.5$ fm. \iffalse The gray line shows the minimum of the vorticity in the region $|x_\bot|<1.5$ fm.\fi The plot below depicts the analogous functions at rapidity $\xi=0.25$. Even at proper times the average vorticity in the central, low rapidity region is only about $2\%$ of the peak vorticity.}}
\label{fig_vort3}
\end{figure}

\begin{figure}[htp]
\begin{center}
\includegraphics[scale=0.65]{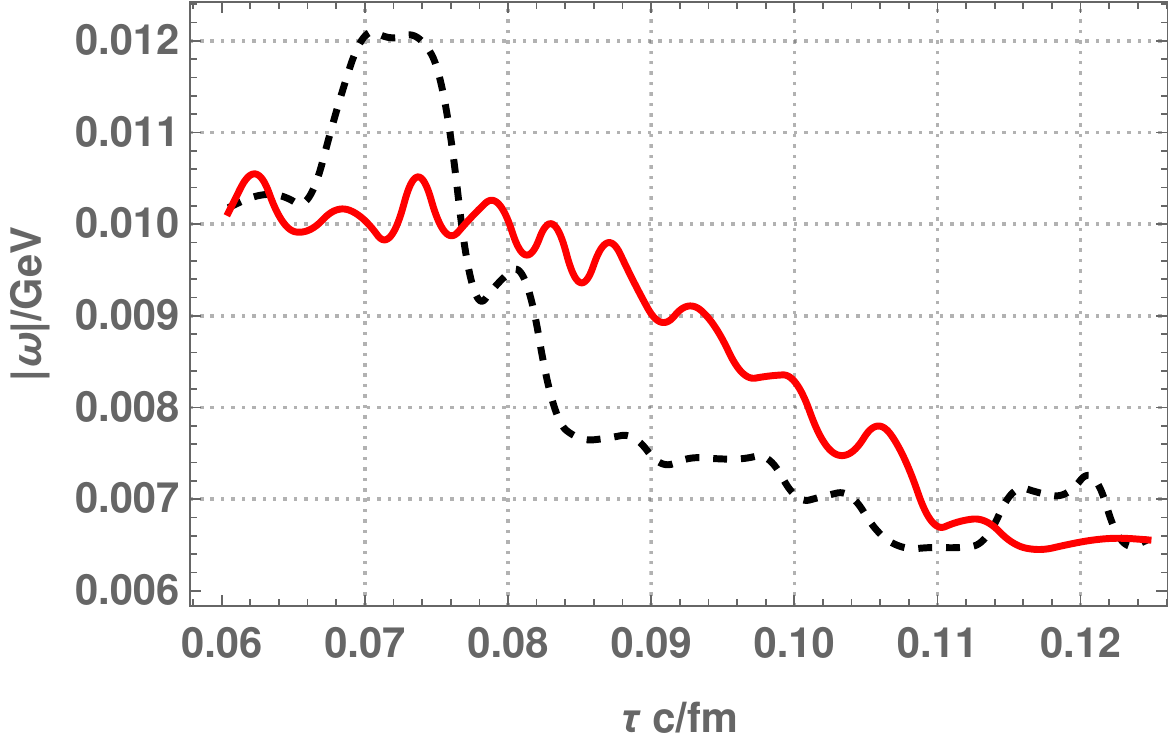}
\end{center}
\vspace{-0.6cm}
\caption{\SWn{The median (red solid curve) and the average (black dashed curve) absolute spatial vorticity $|\vec \omega|$ in the region $\mathcal{R}$ defined in (\ref{R}) as a function of proper time.}}
\label{fig_vort4}
\end{figure}

Examining the vorticity of the produced quark gluon plasma 
is interesting.
There has been much discussion of how the plasma vorticity,
when evolved through to hadronization, may leave signatures
in the polarization of measured $\Lambda$ hyperons
\cite{star}.
Recent advances in
hydrodynamics \cite{spin_hydro}, which now allow one to incorporate
a spin chemical potential into hydrodynamic evolution, open up the
possibility of clarifying to what extent vorticity is responsible
for the observed polarization, starting from the boosted, nuclear
heavy ion model described previously and following the evolution
of vorticity throughout the collision using
holographic modeling of pre-hydrodynamic dynamics followed
by hydrodynamic evolution thereafter.
With this motivation in mind we examine
the vorticity
\begin{equation}
\omega^{\alpha} \equiv
-\frac{1}{2} \, \epsilon^{\alpha \beta \gamma \delta} \,
u_\delta \, \partial_\beta u_\gamma
\label{vort}
\end{equation}
of the early quark gluon plasma at the time when the majority of the central, low rapidity region has  hydrodynamized. In Fig.~\ref{fig_vort} we show the absolute value of the vorticity three vector $|\vec{\omega}|$, with $\omega^\alpha = (\omega^0, \vec \omega)$, at $t=0.1$ fm/$c$.  We find that almost none of the large, initial spatial vorticity $|\vec{\omega}|$ is deposited in the central, hydrodynamized region of the quark gluon plasma. In other words, the plasma is only slowly rotating despite the large, initial  ``geometric'' angular momentum in the system arising from  a large impact parameter.\SW{ In Fig.~\ref{fig_vort3} we show the median vorticity in the central regions  $|x_\bot|<1.5$ fm and  $|x_\bot|<7.5$ fm, the same regions for which we presented the averaged hydro residual $\Delta$ in Fig.~\ref{Delta2}.} \SWn{Likewise, in analogy to Fig.~\ref{Delta3} which shows the hydro residual in the region  $\mathcal{R}$, we depict the average and median vorticity in the region $\mathcal{R}$ in Fig.~\ref{fig_vort4}.}

\begin{figure}
\includegraphics[scale=0.7]{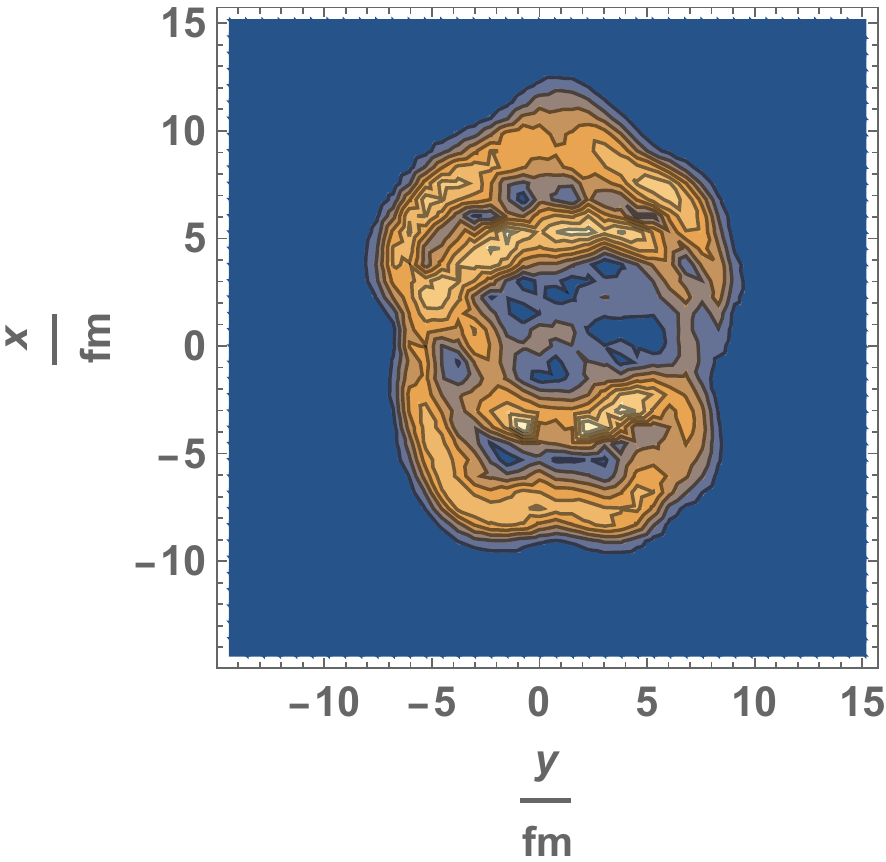}
\includegraphics[scale=0.6]{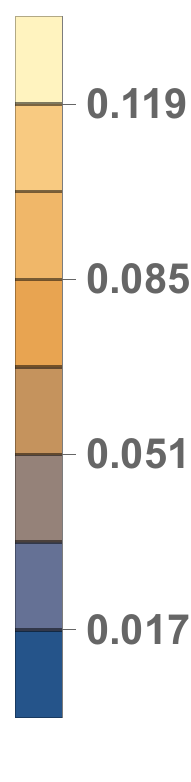}
\includegraphics[scale=0.7]{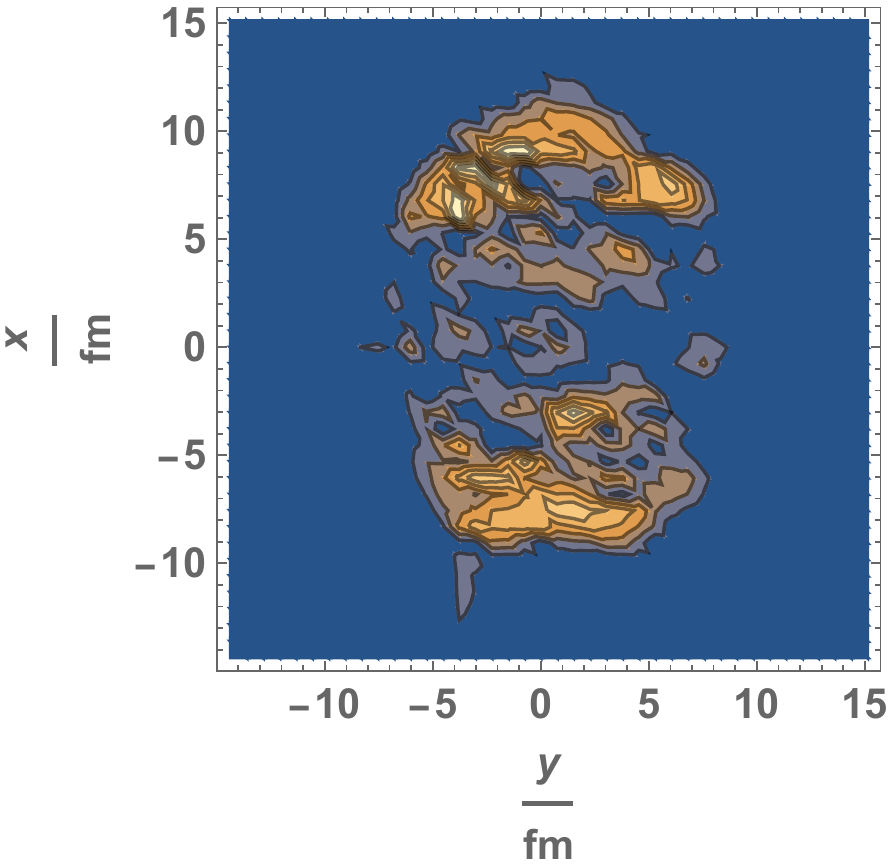}
\includegraphics[scale=0.6]{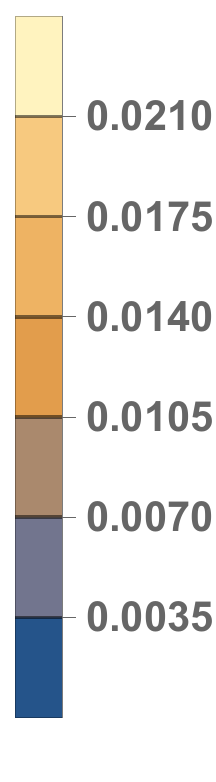}
\includegraphics[scale=0.7]{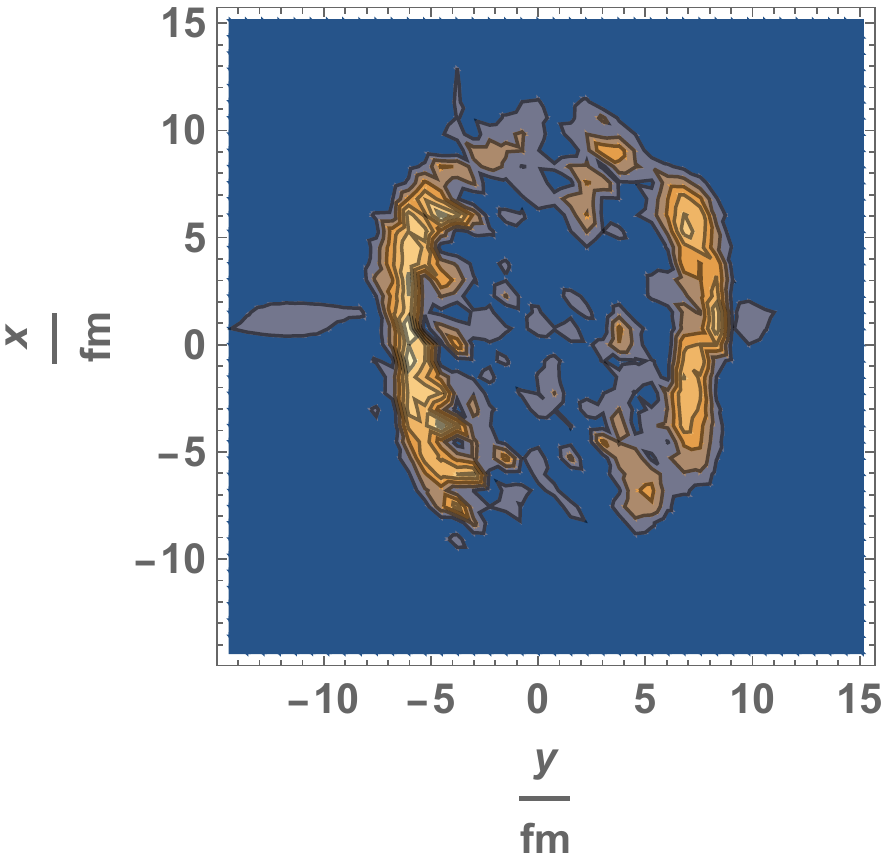}
\includegraphics[scale=0.6]{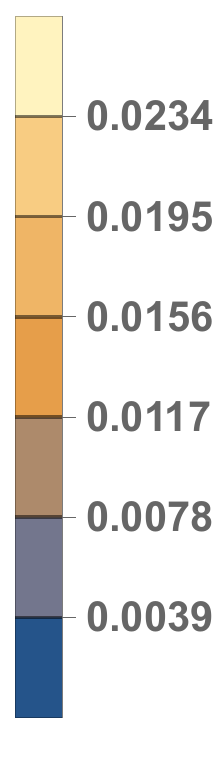}
\includegraphics[scale=0.7]{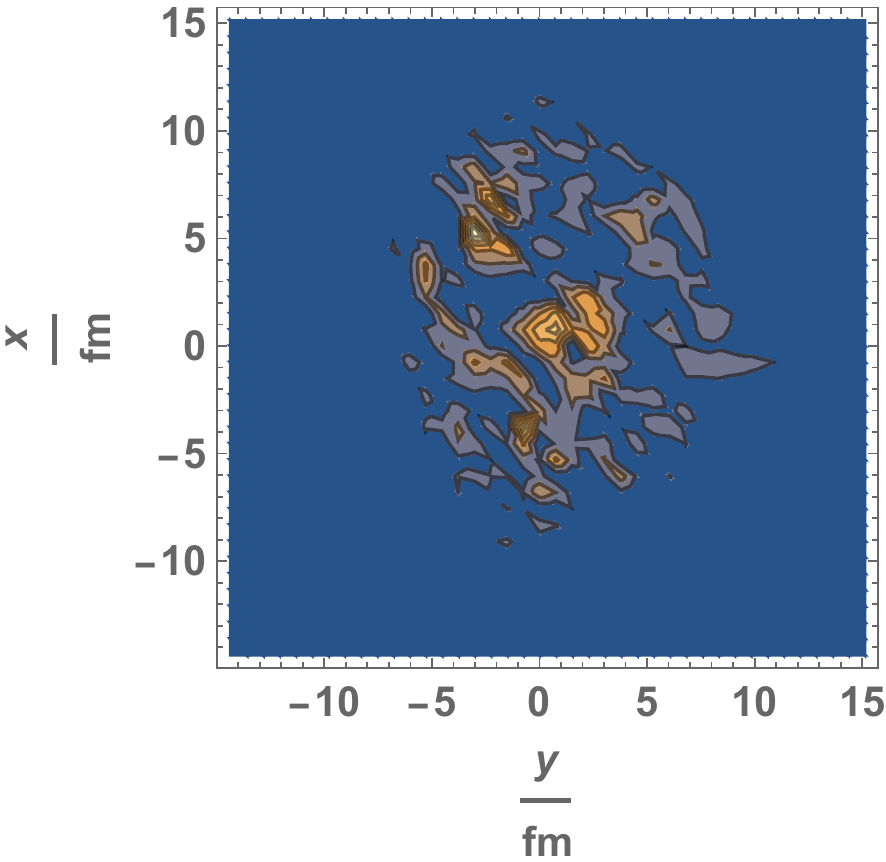}
\includegraphics[scale=0.6]{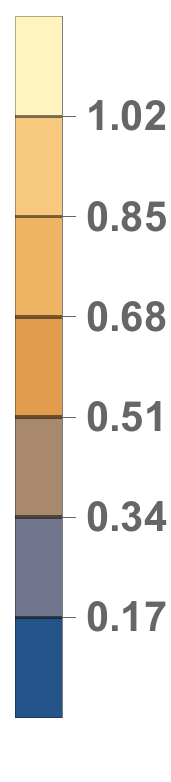}
\vspace{-0.1cm}
\caption{Top left: The absolute value of the spatial components $(\bar \omega^{yz},\bar \omega^{xz},\bar \omega^{xy})$ of the thermal vorticity $\bar \omega^{\mu \nu}$, on the central plane $z=0$. Top right: The absolute value  of $\bar \omega^{tx}$. Bottom left: The absolute value   of $ \bar \omega^{ty}$.  Bottom right: The absolute value of $\bar \omega^{tz} (= -\bar\omega^{zt})$, which is the only non-negligible component of the thermal vorticity at the center of the hydrodynamized part of the plasma. All plots display results at $t=0.1 $ fm/$c$ and are given in units of [GeV].}
\label{fig_vort2}
\end{figure}

 There has also been discussion about the relation between the mean spin vector, and thus the   polarization of emitted spin  $\frac{1}{2}$ particles, and the ``thermal vorticity,'' defined as
\begin{equation}
\bar{\omega}_{\mu \nu} \equiv
\frac{1}{2} \Big(\partial_\mu \beta_\nu -\partial_\mu \beta_\nu \Big),
\label{thermal_vorticity}
\end{equation}
where $\beta^\mu =  u^\mu/T$ with the (local) temperature $T$ inferred from the local energy density.  The authors of  \cite{bec} proposed a relation
\begin{equation}
S^\mu(x,p)\sim (1-n_\text{F})\,\epsilon^{\mu \nu \rho \sigma}\,p_\nu\, \bar \omega_{\rho \sigma}+\mathcal{O}(\bar \omega^2)
\label{spin}
\end{equation}
between the thermal vorticity $\bar\omega$ and the mean spin vector
$S^\mu(x,p)$ and four-momentum $p_\nu$ of an emitted particle
(with $n_\text{F}$ the Fermi-Dirac distribution).
While this relation is, at best,
relevant on the freeze-out surface,
this suggestion motivates us to examine the early development of the
so-defined thermal vorticity.
Fig.~\ref{fig_vort2} shows the size of components of the thermal vorticity $\bar{\omega}^{\mu \nu}$ at the central plane $z=0$.

We find that $\bar{\omega}^{t  z} (=- \bar{\omega}^{z  t})$ is the only component of significant size in the central region at the 
time when the majority of this region has hydrodynamized, $t \approx 0.1 $ fm/$c$. The dominant contribution to $\bar{\omega}_{t  z}$  originates from the time derivative of the longitudinal fluid velocity $\partial_t \, u_z$, which is large compared to transverse components of the fluid velocity. In an idealized setting of perfectly smooth, Gaussian projectiles, the $u_z$ component vanishes at the central point, due to the exact anti-symmetry of $u_z$ in the longitudinal $z$ direction with respect to the origin. However, due to the lumpy structure of our initial data this  no longer holds exactly.

\FloatBarrier

\section{Conclusion}
Solving the Einstein equations, using a truncated expansion in
transverse derivatives, we have numerically calculated, via
gauge/gravity duality, the collision of two highly boosted, lumpy,
localized distributions of energy density in \Nfour\ super Yang-Mills theory.
To model heavy ion collisions,
we craft our gravity initial data to correspond to a state in the
boundary field theory whose stress-energy expectation value
matches a reasonably realistic model of highly boosted and Lorentz
contracted  heavy ions. The parameters we chose reflect those used
in prior modeling of RHIC collisions. This is the first attempt to
use holographic methods to directly investigate the influence of
the nuclear structure of heavy ions on the post-collision flow.
We limited our (real world) computation time to about three weeks,
during which we computed the collision dynamics up until $t=0.144$ fm/$c$
(with $t=0$ corresponding to the time when the longitudinal positions
of the projectiles' centers of mass coincide).
We studied the hydrodynamization time
of the central collision region and found only a modest delay compared
with results that do not incorporate the lumpy structure of the
projectiles: \SW{ In units of the third root of the longitudinally
integrated energy density $\mu$, a  hydrodynamization time of the
low rapidity, central region of $t_{hydro} \approx 1.25 \,\mu$ was
found in \cite{Che}, using analogous units we find that more than half of the   the low rapidity, central region is hydrodynamized at time  $t_{hydro} \approx 1.4
\,\mu$. However, individual transverse plane pixels in the central, low rapidity region are still far from a hydrodynamic description, which is in line with expectations of \cite{mue}.}

We found that the hydrodynamized part of the plasma is only slowly
rotating despite the large, initial, ``geometric'' angular momentum.
The only sizable contributions to the thermal vorticity came from
relativistic corrections. The small vorticity we find early after
the collision in the hydrodynamized region makes it highly unclear
whether subsequent hydrodynamic evolution, up until the
freeze-out surface, will yield a vorticity that is sizable enough
to account for the observed polarization of emitted $\Lambda$ hyperons.

\SW{In future work, we hope to extend the evolution further in time,
long enough to capture the entire hydrodynamization hypersurface
and use the stress energy tensor on this surface as initial data
for subsequent hydro evolutions. Moreover, it will
be very interesting to consider holographic collisions including
a proper treatment of electromagnetism and the spatially distributed
charge and current densities,
and thereby
incorporate the effects of the strong but transient
magnetic background field which develops during heavy ion collisions
and study its effect on the dynamics.
Further future directions include
computing localized collisions including finite coupling corrections
\cite{Gubser,Theissen,Folkestad},
and solving analogous problems in holographic models closer to QCD.}

\section*{Acknowledgments}
  The work of  LY was supported by the U.S. Department of Energy grant DE-SC-0011637. SW acknowledges support by an Israeli
Science Foundation excellence center grant 2289/18 and a Binational Science Foundation grant 2016324. Parts of the work of SW were supported by the U.S. Department of Energy grant DE-SC-0011637 and the Feodor Lynen fellowship program of the Alexander von Humboldt foundation.

\section*{Appendix: Transverse derivative expansion}
\label{transverse_derivatives}
We give a short overview of the approximation scheme, following \cite{2206.01819}, that we used to calculate the holographic collisions. Exploiting the large disparity between longitudinal and transverse scales during heavy ion collisions, we effectively replace transverse derivatives $\partial_{\bot} \rightarrow \epsilon \,  \partial_{\bot}$, expand the Einstein equations in powers of $\epsilon$, solve them order by order, and then set $\epsilon =1$ at the end.

Let us write the Einstein equations for a metric $G$ schematically as 
\begin{equation}
E(G)=0.
\end{equation} 
Expanding in transverse derivatives, we have
\begin{equation}
E(G)=E^{(0)}(G)+ \epsilon \, E^{(1)}(G)+ \epsilon^2\, E^{(2)}(G),
\end{equation}
where the differential operator $E^{(i)}$ contains $i$ powers of transverse derivatives.
   Let $G^{(i)}_{\mu \nu}$ denote an approximate solution to the  Einstein equations  valid to order $\mathcal{O}(\epsilon^i)$ so that \begin{equation}
E(G^{(i)})=\mathcal{O}(\epsilon^{i+1}).
\label{approximate_EQ}
\end{equation} At the lowest order  $G^{(0)}_{\mu \nu}(x^0,x^{||},\bold x^\bot)$ is, for every fixed value of $\bold x^\bot$, some solution to the planar Einstein equations (obtained by neglecting transverse derivatives), with parameters of the specific planar solution varying slowly with $x^\bot$. At zeroth order 
   \begin{equation}
   E^{(0)}(G^{(0)})=0.
   \label{lowest_order_EQ}
\end{equation}    
One now systematically corrects this zeroth order approximation by writing
\begin{equation}
G^{(i)}_{\mu \nu}(x^0,x^{||},\bold x^\bot)=G^{(i-1)}_{\mu \nu}(x^0,x^{||},\bold x^\bot)+\delta g_{\mu \nu}^{(i)}(x^0,x^{||},\bold x^\bot)
\end{equation} 
 and demands that the Einstein equations hold up to the next order.
 Let $\Delta_L^{(i)}$ be the planar Lichnerowicz operator  evaluated on $G^{(i)}$,
\begin{equation}
\Delta_L^{(i)}  \equiv\frac{ \delta E^{(0)} (G^{(i)})}{\delta G^{(i)}}.
\end{equation}
Then Eq. (\ref{approximate_EQ}) will be satisfied if
\begin{equation}
\Delta_L^{(i-1)}  \delta g^{(i)} = - E^{(0)} (G^{(i-1)}) -\epsilon \,E^{(1)}(G^{(i-1)}) -\epsilon^2\,E^{(2)}(G^{(i-2)}).
\label{expanded_EQ}
\end{equation}
See \cite{2206.01819} for a more detailed exposition.

\FloatBarrier
 
\end{document}